\documentclass[journal=jpccck,manuscript=article,layout=twocolumn]{achemso}

\makeatletter
\let\l@addto@macro\relax
\makeatother
\usepackage[fontsize=10pt]{scrextend}
\usepackage{newtxmath,newtxtext}
\usepackage{float}
\restylefloat{table}
\usepackage{chemformula} % Formula subscripts using \ch{}
\usepackage[T1]{fontenc} % Use modern font encodings
\usepackage{multirow}
\usepackage{amsmath}
\DeclareMathOperator*{\argmin}{arg\,min}

\newcommand{\soapt}[0]{\texttt{soap\_turbo}}

\author{Suresh Kondati Natarajan}
\affiliation[Aalto]{Department of Electrical Engineering and Automation,
Aalto University, Espoo 02150, Finland}
\email{suresh0807@gmail.com}
\author{Miguel A. Caro}
\affiliation[Aalto]{Department of Electrical Engineering and Automation,
Aalto University, Espoo 02150, Finland}

\title{Particle Swarm Based Hyper-Parameter Optimization for Machine Learned
Interatomic Potentials}


\begin{document}

\begin{abstract}
Modeling non-empirical and highly flexible interatomic potential energy surfaces
(PES) using machine learning (ML) approaches is becoming popular in molecular and
materials research. Training an ML-PES is typically performed in two stages:
feature extraction and structure-property relationship modeling. The feature
extraction stage transforms atomic positions into a symmetry-invariant
mathematical representation. This representation can be fine tuned by adjusting
on a set of so-called ``hyper-parameters'' (HPs). Subsequently, an ML algorithm
such as neural networks or Gaussian process regression (GPR) is used to model the
structure-PES relationship based on another set of HPs. Choosing optimal values
for the two sets of HPs is critical to ensure the high quality of the resulting
ML-PES model. 

In this paper, we explore HP optimization strategies tailored for ML-PES generation
using a custom-coded parallel particle swarm optimizer (available freely at https://github.com/suresh0807/PPSO.git). We employ the smooth overlap
of atomic positions (SOAP) descriptor in combination with GPR-based Gaussian
approximation potentials (GAP) and optimize HPs for four distinct systems:
a toy C dimer, amorphous carbon, $\alpha$-Fe and small organic molecules (QM9 dataset).
We propose a two-step optimization
strategy in which the HPs related to the feature extraction stage are optimized
first, followed by the optimization of the HPs in the training stage. This strategy
is computationally more efficient than optimizing all HPs at the same time by means
of significantly reducing the number of ML models needed to be trained to obtain
the optimal HPs. This approach can be trivially extended to other combinations of
descriptor and ML algorithm, and brings us another step closer to fully automated
ML-PES generation.
\end{abstract}


\section{Introduction}
Studying the dynamic motion of atoms in molecular and material systems with
femtosecond resolution is made possible via molecular dynamics (MD)
simulations.~\cite{allen2017computer} In the last few decades our reliance on
such simulations to understand complex systems has increased steadily, driven by
advances in computer architectures and the availability of more accurate mathematical
functions describing interatomic interactions.~\cite{hollingsworth2018molecular}
Numerical integration of Newton's equations of motion, one time step at a time,
to sample the phase space of the model system, requires access to the forces acting
on the atoms in the system, i.e., the negative gradient of the potential energy surface (PES).
Historically, these forces are obtained from empirical potentials (``force fields'')
that incorporate physically inspired functional forms fitted to experimental
and/or \textit{ab initio} reference data. This type of simulation is termed
\textit{classical} molecular dynamics (CMD).~\cite{mackerell2004empirical}
As computer power increased, it became possible to compute accurate forces
on-the-fly from the electronic structure of the atomic system at every time step,
termed \textit{ab initio} molecular dynamics (AIMD).~\cite{marx2009ab} There is
an inescapable trade off between accuracy and flexibility, on the one hand, and
computational cost, on the other, to simulate atomic systems. Thus, there is a
great drive to develop new methods that are as fast as CMD and as accurate as AIMD.
This is where PES learned by machine learning algorithms (MLAs) present themselves
as an optimal solution.~\cite{handley2010potential,behler2016perspective,
schmitz2019machine,mueller2020machine,dral2020hierarchical,deringer_2019}

MLAs enable us to understand, find patterns in and predict future outcomes of
complex systems from huge volumes of data accumulated from both experiments
and simulation.~\cite{liu2010advanced,butler2013progress,fenton2018advances,
de2019new,schmidt2019recent,2019-challenges-Bryce,morgan2020opportunities} Establishing
a strong structure-property relationship is at the core of the ML strategy in
developing applications for chemical systems. These ML models rely on a number
of so-called hyper-parameters (HPs) affecting their performance. These
HPs control the learning rate (how much data is needed to achieve a certain
accuracy) and the interpolation power (what accuracy can the ML model achieve
with a given data set). Therefore, HP optimization is a critical step towards
ensuring that a given ML model is making the most out of the available data,
and an integral part in the training workflow, which for ML-PES models can
be summarized as follows:
\begin{enumerate}
    \item Building a reference database of atomic positions and corresponding
    properties (energies, forces, etc.)
    \item Symmetry-invariant feature extraction from atomic positions
    \item Hyper-parameter optimization
    \item Training and validating the ML model
\end{enumerate}

Building a reference database includes careful selection of reference structures
representing the material/chemical system and a reference method to compute their
properties. The reference method is usually more expensive to evaluate than
the ML model. A reference method that is commonly used to describe condensed-phase
systems with reasonable accuracy and felxibility is Kohn-Sham density functional
theory (DFT).~\cite{kohn1965self} However, researchers have
also used more accurate (and relatively more expensive) wave function based methods
as reference to train ML models of molecular systems.~\cite{schran2018high,chmiela2018}

To validate the trained ML model, a test set must be created. This could simply be
a fraction of the reference database (``train/test split'') or could involve a complex
suite of test simulations (stiffness tensor, phonon spectra, phase diagrams, etc.).
Usually, the former is a good choice in the preliminary stages, whereas the most
promising ML models could be further scrutinized using the latter.

Since the atomic positions of a reference structure are not invariant with respect
to rotation and translation of atoms, they are unsuitable to be used directly as
input for training the ML model. Therefore, positions must first be converted to an
invariant representation. This representation could encode the entire structure,
termed as `global descriptor', such as Coulomb matrix~\cite{rupp2012fast} and
many body tensor representation,~\cite{huo2017unified} or every atom in the
system individually, termed as `local descriptor'.~\cite{himanen2020dscribe}
For training ML-PES of condensed-phase systems, local descriptors are desirable
and the total energy of the system can be computed as a sum of atom-wise energy
contributions.~\cite{christensen2020fchl} This also allows for transferability of
the ML-PES. There are several descriptor commonly used in the representation of
atomic structures.~\cite{himanen2020dscribe} Two of the most widely used approaches
are the smooth overlap of atomic positions
(SOAP)~\cite{bartok2013representing,caro2019optimizing} and atom-centered symmetry
functions (ACSF).~\cite{behler2011atom} These approaches encode the immediate
chemical environment of every atom in the system into a set of invariants
(simply `descriptors', from here onward), which form the training data along
with the properties obtained from the reference method. It has been shown that
these representations are formally related to one another~\cite{willatt2019atom}.
Imbalzano \textit{et al}.~\cite{imbalzano2018automatic} have previously identified
efficient methods to automatically optimize the size of ACSF and SOAP fingerprints
to describe the atomic environment of the desired chemical system. We also note
there exist neural networks that use deep learning to extract the symmetry invariants
directly from the input 3D structural data.~\cite{schutt2017quantum} For ML-PES training,
the reference properties typically include the total energy of the structure, atomic
force components, virial stress components and atomic charges. 

The ML algorithm is responsible for establishing the structure-property relationship.
The two most popular ML algorithms for condensed-phase ML-PES training are
Behler-Parrinello type neural networks (BPNN)~\cite{behler2007generalized} and
Gaussian process regression (GPR, generally known in the community as Gaussian
approximation potential, GAP).~\cite{bartok2010gaussian,bartok2015g} GPR is a particular
flavor of the more general class of kernel-based regression
algorithms~\cite{christensen2020fchl}. Both ML
and descriptor algorithms have adjustable HPs, that need to be optimized for the
target system. Since training an ML model is an expensive endeavor in terms of
computational time, rigorous HP optimization (HPO) is often impractical due to
the large HP search space. For the same reason, the HPs in the earlier ML-PES
models have been selected either from chemical intuition~\cite{bernstein2019novo},
grid search, stochastic search,~\cite{christensen2020fchl}
or from testing a small number of HP combinations using approaches similar to
the design of experiments. More recently, we performed
a Sobol sequence based HP search to obtain optimal HPs for an amorphous carbon GAP.\cite{caro2019optimizing}
Schmitz \textit{et al.}~\cite{schmitz2019machine} performed HP optimization of a GPR
model of the \ch{F2} molecule using open source optimizers. Recently, an application
to automate molecular PES training using the HyperOpt~\cite{bergstra2015hyperopt}
optimizer has been introduced.~\cite{doi:10.1021/acs.jctc.9b00312} It is important
to note that an accurate model of the whole HP surface is not needed because we
are only interested in its global minimum.

While there are deterministic algorithms that are guaranteed to find the global
minimum of high dimensional functions, they are impractical for finding solutions to
multi-parameter dependent real world problems, which are in most cases NP-hard.
By contrast, heuristics-based stochastic algorithms, although not guaranteed to
find the global minimum in every instance, have the best chance to find solutions
to such problems.~\cite{collet2008stochastic} There are several non-traditional HPO
techniques for ML models available in the literature such as Bayesian optimization,
random forest algorithm and particle swarm
optimization~\cite{bergstra2015hyperopt,Feurer2019}. However, we are unaware of any
HPO strategy tailored specifically for complex ML-PES training. A specialized
strategy is necessary since the impact of each hyper-parameter on the accuracy and
performance of the ML-PES model is not the same. While good values for
hyper-parameters can be guessed for simple and well studied chemical systems,
guessing them is impossible for new and complex systems. In this paper, we look
at optimization strategies specific to the HPs encountered in ML-PES training
using a custom coded parallel particle swarm optimizer, which is a heuristic based
stochastic optimizer. We will consider the combination of SOAP descriptors and
GAP ML model in this paper, but it is possible to trivially extend the discussed
strategies to other descriptors and ML schemes as well. Incorporating these
strategies into the training workflow will bring us one step closer to efficient
fully automated ML-PES generation.

%%%%%%%%%%%%%%%%%%%%%%%%%%%%%%%%%%%%%%%%%%%%%%%%%%%%%%%%%%%%%%%%%%%%%%%%%%%%%%%%%%%%%%
\section{Hyper-Parameter Classification}
\label{sec:HPclass}
\begin{figure*}
    \centering
    \includegraphics[scale=0.65]{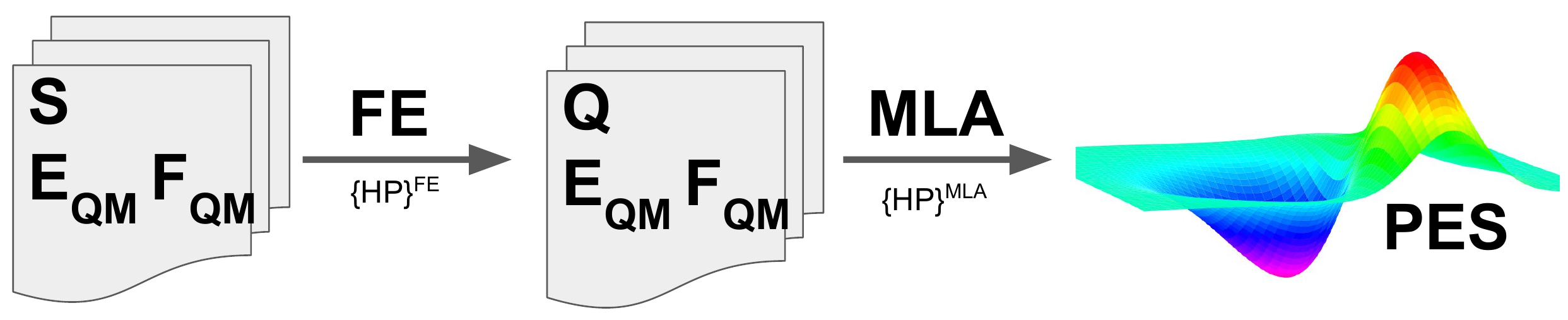}
    \caption{A schematic representation of the steps involved in the ML-PES training.
    FE and MLA are acronyms for `feature extraction' and  `machine learning algorithm'
    and they have hyper-parameters of their own, \{HP\}$^\text{FE}$ and
    \{HP\}$^\text{MLA}$. $S$ is the 3D structure data, $Q$ is the set of descriptors
    representing $S$, $E^\text{QM}$ and $F^\text{QM}$ are the quantum mechanical
    energies and forces of $S$. \{$Q$, $E^\text{QM}$ and $F^\text{QM}$\} forms the
    reference data set that will be used to train the ML-PES model.}
    \label{fig:hpclass}
\end{figure*}

Figure~\ref{fig:hpclass} shows an overview of the steps involved in a ML-PES
training workflow. There are two important stages, the feature extraction (FE) stage and the machine learning algorithm (MLA) stage, in which two distinct sets of HPs are needed. The FE stage converts the atomic coordinates $S \equiv \{ \textbf{r}_i \}$ in the reference database into descriptors $Q \equiv \{ \textbf{q}_i \}$. This transformation depends parametrically on a set of HPs, \{HP\}$^{\rm{FE}}$. The MLA stage relates the descriptors $Q$ to
the corresponding quantum chemical reference data ($E^{\rm{QM}}$ and $F^{\rm{QM}}$),
based on another set of HPs, \{HP\}$^{\rm{MLA}}$.

In the FE stage, \{HP\}$^{\rm{FE}}$ controls how the representation of the atoms
is carried out. For
many-body descriptors, such as SOAP and ACSFs, this representation may involve
the assignment of each atomic position to an atomic density, as well as
the construction of a basis set for finite numerical expansion of this atomic
density. This is conceptually very similar to how electron densities are
represented in DFT codes (e.g., by using localized atomic orbitals). The
symmetry-invariant representation of $S$ into $Q$ leads to a mapping from an
$N_\text{at} \times 3$ dimensional space to an $N_\text{at} \times N^{\rm{FE}}$
dimensional space, where $N_\text{at}$ is the number of atoms and
$N^{\rm{FE}}$ is the dimension of each atomic descriptor $\textbf{q}_i$.
This generally leads to a significant increase in dimensionality, since usually
$N^{\rm{FE}} >> 3$. The
\{HP\}$^{\rm{FE}}$ can be further split into two subsets, \{HP\}$^{\rm{FE}}_{\rm{P}}$
and \{HP\}$^{\rm{FE}}_{\rm{S}}$, where \{HP\}$^{\rm{FE}}_{\rm{P}}$ controls the size
($N^{\rm{FE}}$) of the descriptor and the computational effort needed, via
adjusting the number of basis functions, while
\{HP\}$^{\rm{FE}}_{\rm{S}}$ affects the shape of the atomic density
representation (i.e., how a \textit{point} atom is ``smeared'' into a density
\textit{field}). An optimal \{HP\}$^{\rm{FE}}$ is desirable, such that
the resulting descriptors are not only optimally sized so that they will not
be too expensive to compute but also maximally detailed so as to better
differentiate between atomic environments.

In the MLA stage, the quality of the structure-property relationship depends on
the choice of \{HP\}$^{\rm{MLA}}$ along with the choice of \{HP\}$^{\rm{FE}}$
in the FE stage. Similar to the FE stage, \{HP\}$^{\rm{MLA}}$ can also be split
into two subsets \{HP\}$^{\rm{MLA}}_{\rm{P}}$ and \{HP\}$^{\rm{MLA}}_{\rm{S}}$.
For example, in artificial neural networks, a HP$^{\rm{MLA}}_{\rm{P}}$ could be
the number of hidden layer neurons in the network while a HP$^{\rm{MLA}}_{\rm{S}}$
could be the choice of activation function in the hidden layer neurons (e.g., using
a sigmoid function instead of a hyperbolic tangent does not increase the
computational cost but only changes the output value range). The total number of
HPs combining those from FE and MLA stages, $N_{\rm{HP}}$, can be small or large
depending on the choice of the descriptor and MLA. Each point in this $N_{\rm{HP}}$
dimensional space has a ML-PES and a quality estimate associated to it. As
mentioned earlier, our aim is not to model this $N_{\rm{HP}}$ dimensional
objective function but to find its global minimum as cheaply and efficiently
as possible. For the remainder of this paper we will use SOAP in the FE stage
and GAP in the MLA stage.

%%%%%%%%%%%%%%%%%%%%%%%%%%%%%%%%%%%%%%%%%%%%%%%%%%%%%%%%%%%%%%%%%%%%%%%%%%%%%%%%%%%%%
\section{SOAP, GAP and Objective Functions}
Let us first look at how SOAP and GAP allow us to predict energies and forces in a
simple practical way. For more extensive details please refer to the
literature.~\cite{bartok2010gaussian,bartok2013representing,bartok2015g,caro2019optimizing}
Let us consider a training database containing $N_\text{st}$ structures each with
$N_\text{at}$ atoms (or, equivalently, atomic environments). First, the $3 \times
(N_\text{at} \times N_\text{st})$ bits of Cartesian positions information ($S$)
of the reference database are transformed into $N^{\rm{SOAP}} \times (N_\text{at}
\times N_\text{st})$ bits of information ($Q$) based on the choice of
\{HP\}$^{\rm{SOAP}}$:
\begin{align}
    \text{FE} (S; \{\text{HP}\}^{\text{SOAP}}) \longrightarrow Q,
\end{align}
using the SOAP formalism introduced by Bart\'ok
\textit{et al}.~\cite{bartok2013representing} Here, we work with the SOAP-type
descriptors introduced by us elsewhere,~\cite{caro2019optimizing} to which we refer as
\soapt.

Every atomic environment is bounded by a SOAP cutoff sphere (CS) with radius
$r_\text{cut}$. The atomic density of the environment of atom $i$ is approximated by an
expansion in radial basis $g_{n}$ and spherical harmonics $Y_{lm}$:~\cite{bartok2013representing}
\begin{align}
    \rho_\text{env} (i) = \sum_{j \in \text{CS}_i} \sum_{nlm} c^j_{nlm} (i) \, g_{n}(r_j)
    Y_{lm}(\theta_j,\phi_j).
\end{align}
Here, $n$ and $l$ are the indices for radial and angular channels, respectively. The maximum
values of $n$ and $l$ control the size of the basis. $m$ takes the integer values
between $-l$ and $l$. The outer
sum is over all the atoms, $j$, that are within the cutoff sphere of $i$ (CS$_i$),
and the $c_{nlm}$ are the expansion coefficients. The $N^\text{SOAP}$ components of
the SOAP descriptor $\textbf{q}_i$ for atomic environment $i$ are obtained from the normalized
power spectrum $\textbf{p}_i$ of the corresponding atomic density:
\begin{align}
    & p_{nn'l} (i) = \sum_{m=-l}^{l} [c_{nlm}(i)]^* c_{n'lm} (i),
    \nonumber \\
    & c_{nlm} (i) = \sum_{j \in \text{CS}_i} c^j_{nlm} (i),
    \qquad
    q_{nn'l} (i) = \frac{p_{nn'l} (i)}{\sqrt{\textbf{p}_i \cdot \textbf{p}_i}}.
\end{align}

The total number of invariant $N^\text{SOAP}$ components describing the atomic environment,
with $N_\text{sp}$ elemental species in it, is given by
\begin{align}
    N^\text{SOAP} = \frac{n_\text{max}^2 N_\text{sp}^2 + n_\text{max} N_\text{sp}}{2}
    \left(l_\text{max} + 1\right),
\end{align}
where $n_\text{max}$ and $l_\text{max}+1$ are the total number of radial and angular
momentum channels used in the basis set, respectively (the division by 2 is for the
symmetry of the coefficients upon exchange of the radial basis
indices;~\cite{caro2019optimizing} $+1$ next to $l_\text{max}$ is to
include $l=0$). To get an idea of the typical dimensionality of SOAP descriptors, consider
the following example: if we choose $n_\text{max}=10$ and $l_\text{max}=10$, we end up with
605 SOAP components for a single species descriptor ($N_\text{sp}=1$). This value
increases to 2310 components for $N_\text{sp}=2$. Therefore, computing SOAP descriptors
becomes increasingly more expensive as the atomic environment becomes more heterogeneous
(``curse of dimensionality'').
Hence, $n_\text{max}$ and $l_\text{max}$ are critical performance-based HPs. Another HP
that significantly affects the performance is $r_\text{cut}$, which determines the total
number of neighbor atoms to be considered inside the SOAP sphere for the many-body
descriptor evaluation (the cost of building SOAP descriptors and descriptor gradients
grows as $r_\text{cut}^3$). Brief descriptions of all \{HP\}$^{\rm{SOAP+GAP}}$ are given
in Table~\ref{tab:desc}. From Table~\ref{tab:desc}, there are 11 {\soapt} HPs in total out of
which 3 are performance affecting HPs and the other 8 are shape affecting HPs. Except
$l_\text{max}$, all the other HPs can be given as a vector of $N_\text{sp}$
components. Choosing different values for each species increases the total number of
unique SOAP HPs to $10 \times N_\text{sp}+1$. We also note that this is the
\textit{maximum} number of HPs to be optimized; one can choose to constrain the search
to those HPs that are expected to have the biggest impact on model accuracy, and set
the rest to the default values.

\begin{table*}[ht]
    \centering
    \begin{tabular}{lccl}
    \hline
          HP     & Symbol        & Units & Description  \\
         \hline\\
         \multicolumn{2}{c}{\{HP\}$^{\rm{SOAP}}_{\rm{P}}$}\\
                       $n_\text{max}$ & $n_\text{max}$ & & number of radial channels \\
                       $l_\text{max}$ & $l_\text{max}$ & & number of angular momentum channels\\
                       rcut\_hard     &rc$_{\rm{h}}$ & \AA & cutoff radius for SOAP sphere\\\\
        \multicolumn{2}{c}{\{HP\}$^{\rm{SOAP}}_{\rm{S}}$}\\              
                       rcut\_soft     & rc$_{\rm{s}}$ & \AA & density smoothing enabled beyond this value\\
                       atom\_sigma\_r & $\sigma_r$ & \AA & width of Gaussian functions in radial channels\\
                       atom\_sigma\_t & $\sigma_t$ & \AA & width of Gaussian functions in angular channels\\
                       atom\_sigma\_r\_scaling & a$_r$ & & radial dependence of functions in radial channels\\
                       atom\_sigma\_t\_scaling & a$_t$ & & radial dependence of functions in angular channels\\
                       amplitude\_scaling & a & & radial dependence of neighbour contribution\\
                       central\_weight & cw & & Gaussian width for the central atom\\
                       radial\_enhancement & re & & factor to enhance the amplitude of radial Gaussians\\\\
                       \hline\\
                       zeta & $\zeta$ & & SOAP kernel is raised to the power of $\zeta$\\\\
                      \hline\\
        \multicolumn{2}{c}{\{HP\}$^{\rm{GAP}}_{\rm{P}}$}\\ 
                        n\_sparse & $n_{\text{sparse}}$ & & number of sparse configuration to be used\\\\
         \multicolumn{2}{c}{\{HP\}$^{\rm{GAP}}_{\rm{S}}$}\\ 
                        delta & $\delta$ & eV & energy contribution scaling factor\\
                        sigma\_E & $\sigma_E$ & eV & regularization parameter for energies\\
                        sigma\_F & $\sigma_F$ & eV/\AA & regularization parameter for forces\\\\
                        \hline
    \end{tabular}
    \caption{Description of SOAP and GAP hyper parameters.}
    \label{tab:desc}
\end{table*}

In the next step, a GAP is generated by Gaussian process regression (GPR) from the reference
energies and forces, where the covariance between two atomic energies $\epsilon_i$ and
$\epsilon_j$ is obtained from the SOAP-based
kernel (which measures the similarity between atomic environments $i$ and $j$):
\begin{align}
    \text{cov} \left( \epsilon_i, \epsilon_j \right)
    := \delta^2 k( \textbf{q}_i, \textbf{q}_j ).
\end{align}
The MLA task depends on the observables ($E^\text{QM}$ and $F^\text{QM}$) and descriptors
($Q$) and, parametrically, on \{HP\}$^{\rm{GAP}}$:
\begin{align}
    \rm{MLA}(Q, E^\text{QM}, F^\text{QM}; \{\rm{HP}\}^{\rm{GAP}})
    \longrightarrow \{\alpha_t\}.
\end{align}
As a result, we produce a set of fitting coefficients $\{\alpha_t\}$, one coefficient for each
training environment, which will be used to predict energies and forces of new geometries. A
detailed account of this procedure is given by Bart\'ok and Cs\'anyi.~\cite{bartok2015g}.
In GAP, the total energy of a structure is predicted as a sum of atomic environment based
energy contributions:
\begin{align}
    E_\text{total}^\text{GAP} = \sum_{i=1}^{N_\text{at}} \epsilon_i^\text{GAP}
    \\
    \epsilon_i^\text{GAP} = \delta^2 \sum_{t=1}^{N_\text{t}} \alpha_t
    k( \textbf{q}_t, \textbf{q}_{i} ).
    \label{eq:local_energy}
\end{align}
Here, the SOAP kernel compares the set of SOAP vectors describing the $N_\text{t}$ training
configurations ($Q_\text{t} \equiv \{ \textbf{q}_t \}$) and $\textbf{q}_i$, the SOAP vector
describing the $i^\text{th}$ atomic environment in the predicted geometry. This kernel outputs
a vector of similarity measures, each component between 0 and 1, by comparing $\textbf{q}_{i}$
with all the vectors in $Q_\text{t}$. A similarity measure of 1 means the two compared environments
are identical with respect to rotation and translation of atoms whereas a value of 0 means they
are nothing alike. SOAP descriptors are usually used in combination with a dot product kernel,
\begin{align}
    k( \textbf{q}_t, \textbf{q}_{i} ) = \left( \textbf{q}_t \cdot \textbf{q}_{i} \right)^\zeta,
\end{align}
where $\zeta$ is a SOAP kernel parameter (with $\zeta >1$ down scaling the similarity
measure, which helps towards better differentiation of similar environments). Carrying out
the element-wise dot products between $Q_\text{t}$ and $\textbf{q}_i$ (and rising each
element to the power of $\zeta$) then outputs a vector of kernels which, when multiplied with
the vector $\{\alpha_t\}$, gives the scalar energy contribution of atom $i$ ($\epsilon_i$)
to the total energy. Forces are then evaluated from the gradients of these energy contributions
with respect to the Cartesian coordinates of the atoms, which is more demanding due to the
inclusion of the descriptor derivatives and, especially, due to the explicit dependence on the
number of atomic neighbours. $\delta$ represents a scaling factor corresponding to the
distribution of energy contributions. While for single-kernel GPR an optimal
$\delta$ can be obtained in closed form,~\cite{cawley2005estimating} in the context of GAP
$\delta$ is in practice another HP. If multiple GAPs are involved, there will be a
$\delta$ associated with each GAP,
which indicates the distribution of that GAP's contribution to the total energy.
It has been shown that explicitly including two-body and three-body terms improves the overall
quality of the PES models.~\cite{deringer2017machine} However, in this paper we consider GAP
training with SOAP as the only descriptor. The methodology discussed here can be trivially
extended to include two-body and three-body GAPs. $\sigma_E$ and $\sigma_F$ are the
regularization parameters that indicates the expected error in the input energies and forces,
respectively.

The quality of the GAP can be evaluated as root mean squared energy ($\Gamma_{\rm{E}}$)
and force component errors ($\Gamma_{\rm{F}}$),
\begin{align}
    \Gamma_E = \sqrt{\sum_{i=1}^{N_\text{st}}(E^\text{GAP}_i - E^\text{DFT}_i)^2/N_\text{st}}
    \\
    \Gamma_F= \sqrt{{\sum_{i=1}^{3 \times N_\text{st} \times N_\text{at}}(F^\text{GAP}_i
    - F^\text{DFT}_i)^2}/{(3 \times N_\text{st} \times N_\text{at})}}.
\end{align}
$\Gamma_{E}$ and $\Gamma_{F}$ will vary as we change \{\rm{HP}\} for a given $S$,
$E^\text{QM}$ and $F^\text{QM}$:
\begin{align}
    \Phi (\{\text{HP}\} | S, E^\text{QM}, F^\text{QM}) \longrightarrow
    \Gamma_E, \Gamma_F.
\end{align}
Here, $\Phi$ is an unknown black box function in the $N_\text{HP}$ dimensional
space whose global minimum will give the least possible energy and force errors
for the GAP. So, $\Phi$ is the objective function to be minimized:
\begin{align}
    \{\text{HP}\}^\text{opt} =  \argmin_{\{\text{HP}\}} \left( \Phi (\{\text{HP}\} |
    S, E^\text{QM}, F^\text{QM}) \right).
\end{align}
Note that even a \textit{single} evaluation of $\Phi$ can be rather expensive, since it
involves both training and validating a GAP. At the same time, $\Phi$ may have
many local minima. These two facts make the use of traditional optimization methods
that rely on derivatives of the objective function (e.g., quasi-Newton methods)
both expensive and ineffective at finding the global minimum. Hence the need for
efficient global optimization methods like the one presented here.

The number of HPs in the GAP stage is relatively small when compared to that of the
SOAP stage. The only performance affecting HP here is the number of sparse configuration
representing the entire database of atomic environments. When a number smaller than the
total number of environments in the database is chosen, then Eq.~(\ref{eq:local_energy})
runs over the $N_\text{sparse}$ configurations in the sparse set, rather than over
every entry in the training data base.~\cite{bartok2015g}

Therefore, minimizing $n_\text{sparse}$ gives cost benefit in the training as well as
in the prediction stages. The selection of the sparse configurations needs to be done
carefully, since it affects the accuracy of the potential (e.g., if all the sparse
configurations are similar, the potential will generalize poorly). For diverse and
well-balanced data bases, random selection of sparse configurations may be sufficient.
Generally, one may want to pick the $n_\text{sparse}$ most representative configurations,
for instance using $CUR$ matrix decomposition, data clustering techniques and so
on.~\cite{bartok2015g} Optimal reference configuration selection is far from trivial
and remains an active area of research.

%%%%%%%%%%%%%%%%%%%%%%%%%%%%%%%%%%%%%%%%%%%%%%%%%%%%%%%%%%%%%%%%%%%%%%%%%%%%%%%%%%%%%%
\section{Methodologies for automatic HP optimization}
\label{sec:Method}
\begin{figure*}
    \centering
    \includegraphics[scale=0.9]{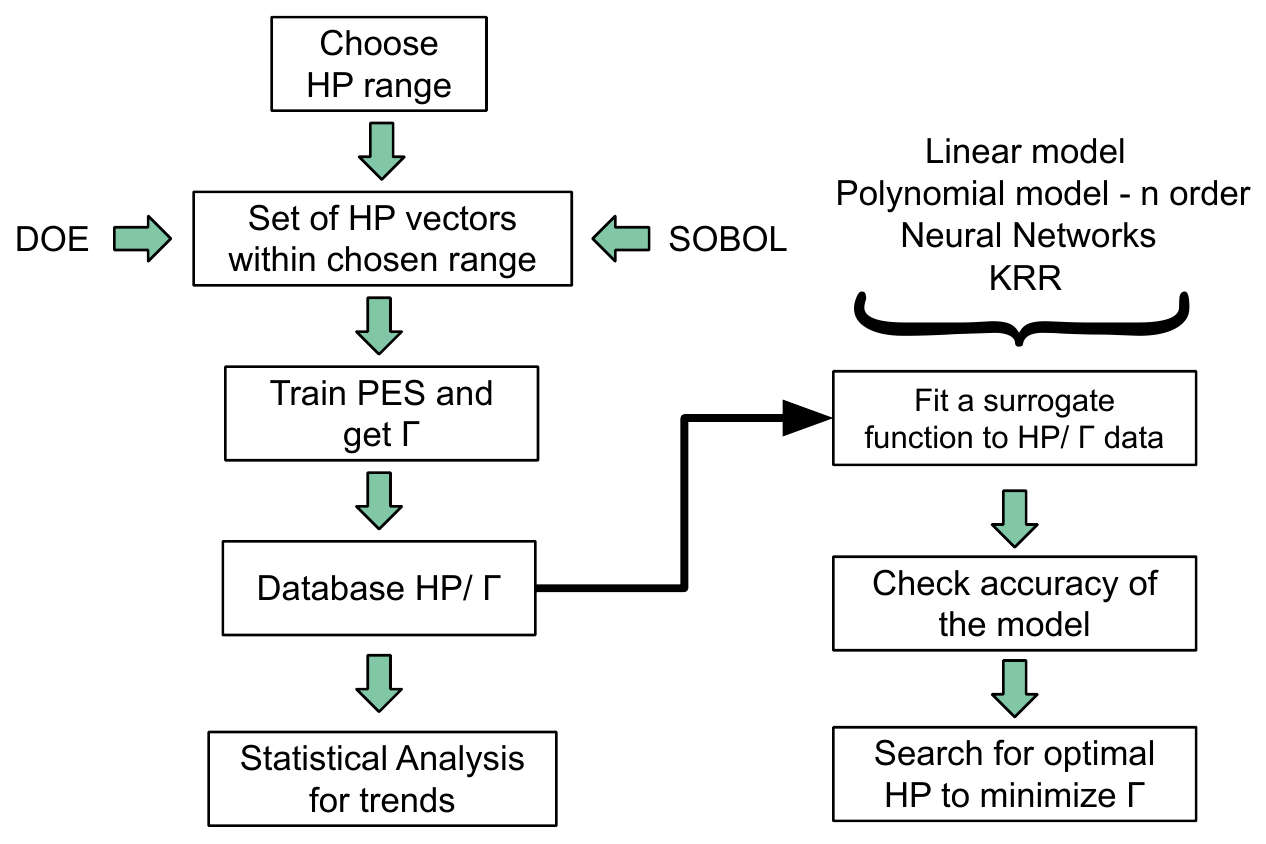}
    \caption{A flowchart of steps involved in offline optimization of HPs.}
    \label{fig:offline}
\end{figure*}

There are two broad categories under which automatic HPO can be carried out:
offline and inline optimization. A flowchart showing the steps involved in an offline
optimization is given in Figure~\ref{fig:offline}.
In offline optimization, a cheap surrogate function (response surface) is modelled to
represent $\Phi$, the $N_\text{HP}$ dimensional objective function, followed by a
global minimum search to find the optimal HPs. This is needed when the cost of evaluating
the objective function is very high. To begin with, a working range for the HPs is chosen
and a number of HP vectors are chosen based on techniques such as the design of experiments
(DOE) approach~\cite{anderson2000design} using orthogonal arrays~\cite{hedayat2012orthogonal}
or Sobol sequences.~\cite{SOBOL196786} For each of the chosen HP vectors, a ML-PES
is trained and the corresponding response parameter ($\Gamma$, RMSE) is recorded in a
database. The database can be directly analysed to understand the influence of each HP on
the response variable. This is an advantage of offline optimization methodology mainly
because of how the HP vectors are sampled. This database is then used to build a surrogate
function using standard techniques such as a linear model or an $n$-order polynomial model.
If the coefficient of regression ($R^2$) is not suitable, then one could train a NN model
on the data. Once a reliable model of the HP surface is ready, it is trivial to search for
minima using standard tools. All the steps involved in offline optimization using DOE data
can be performed in Python.~\cite{sureshDOE}

\begin{figure}[h]
    \centering
    \includegraphics[scale=0.9]{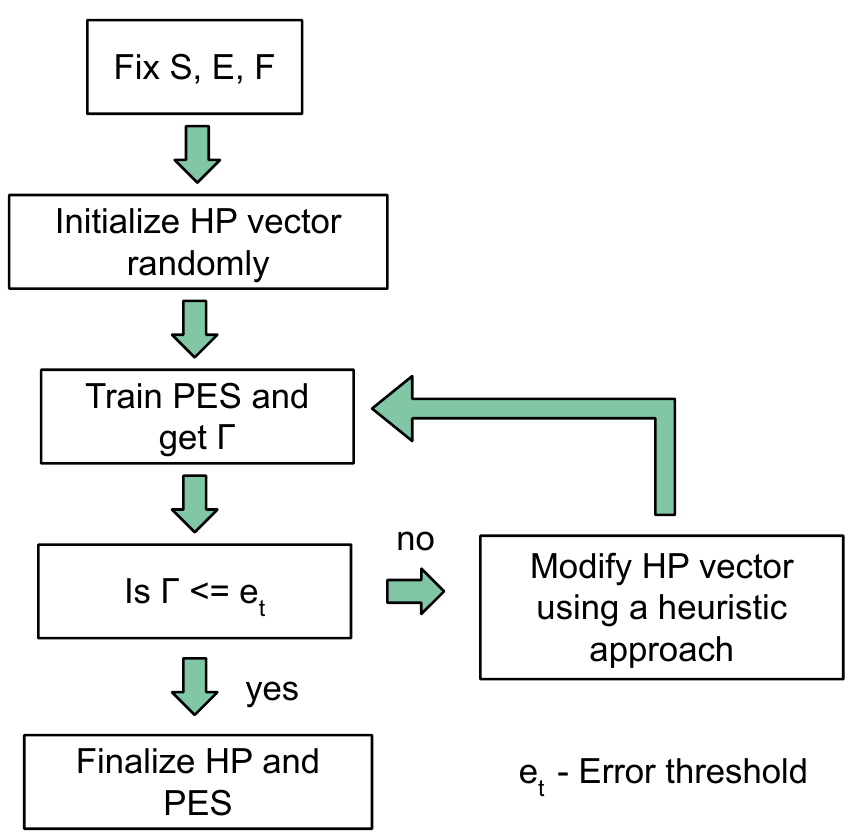}
    \caption{A flowchart of steps involved in inline optimization of HPs.}
    \label{fig:inline}
\end{figure}

A simple flowchart of inline optimization is given in Figure~\ref{fig:inline}.
In inline optimization, a HP vector is chosen randomly and a preliminary ML-PES is
trained and the response parameter ($\Gamma$, RMSE) is computed. We either set a
threshold value $e_\text{t}$ for the response parameter, or a maximum number of
steps can be set alternatively. If $\Gamma$ is less than $e_\text{t}$, then the
optimal HP vector and corresponding ML-PES are obtained. If not, the HP vector is
modified based on a heuristic algorithm and another ML-PES is trained with this
modified HP set. This process is repeated until $\Gamma$ is below the threshold
or the maximum number of steps are reached. This is the preferred approach if the
PES training is not very expensive.

A popular HPO technique used in the ML community is Bayesian optimization
(B-opt),~\cite{snoek2012practical} which is a hybrid of the inline and offline
methodologies discussed above. B-opt requires a starting database of HP vectors
and corresponding $\Gamma$ from which a surrogate function is built using GPR,
similar to the offline optimization. A small number of samples compared to the offline
optimization is enough to start B-opt. The surrogate function, which is also the
objective function here, is defined exactly at the prior data points in the HP space
while the estimated error in $\Gamma$ at other regions is given by a probability
distribution function. An acquisition function is then generated based on the surrogate
function and the associated error probabilities. This acquisition function shows us
the most promising regions in the HP space to sample the next points. This aspect is
similar to the inline optimization. 

If the ML-PES training is costly and the HP space is huge, then all of the above
techniques will be very time consuming. The classification schemes of the HPs discussed
in the earlier section allow us to divide and conquer the HP space, but it will not
help us reduce the cost of training ML-PESs. One solution to this problem is to
train ML-PESs only using reference energies (i.e., leaving forces out) but evaluate
them using both response parameters, $\Gamma_E$ and $\Gamma_F$. It is to be noted that
there will be regions in the HP space where one of the response parameters is minimal
while the other is large, which are not optimal. By directing the search towards the
regions in the HP space where both energy and force errors are minimal, we can obtain
optimal HPs without explicitly training the models with forces.

Once optimal HPs are found, we can further improve the quality of ML-PES by refining the
HP search, retraining the GAPs including forces. The rationale for leaving forces out
of the fit in a first step is that there are a lot more reference forces than reference
energies for a typical atomic structure data base. In particular, the number of reference
energies equals the number of structures (i.e., ``simulation boxes'') in the data base
$N_\text{st}$, whereas the number of forces equals $3 N_\text{st}$ times the average
number of atoms in a structure. For a typical data base, the number of forces is
therefore between one and two orders of magnitude larger than the number of energies. For
the practical purposes of training a GAP, this means increasing CPU times and memory usage
by the same relative amount.

Since GAPs can be trained cheaply by not including forces, we will use inline
optimization to optimize our HPs. For this we will use a parallel particle swarm
optimization (PPSO) algorithm, a heuristic based stochastic algorithm. An in house
MPI Fortran implementation of the PPSO algorithm is used in this work. Our code is
available free of charge at \url{https://github.com/suresh0807/PPSO.git}. There are
4 general steps in the PSO algorithm as follows:

\noindent 1) Choose the number of particles in the swarm and initialize their positions
in the user defined $N_\text{HP}$ dimensional space and initialize their velocities
to 0. Each HP coordinate ($x^i, \, i=1,\cdots,N_\text{HP}$) of a particle is chosen
according to
\begin{align}
    x^{i} (t_0) = x^{i}_\text{min} + R (x^{i}_\text{max} - x^{i}_\text{min}).
\end{align}
Here, index $i$ runs from 1 through $N_\text{HP}$ dimensions of the search space
and $R$ is a random number between 0 and 1, which allows for an unbiased selection
of the particle coordinates. $x^{i}_\text{min}$ and $x^\text{i}_\text{max}$ are the
lower and upper bounds of the $i$th HP, respectively.

\noindent 2) For each particle (HP vector), train a GAP and compute the response
variable(s). At $t=0$, the initialized HP coordinates of each particle are labeled
as the particle's best location or simply local best ($\textbf{x}_\text{l}$). From
the swarm, the coordinates of the particle with the lowest response value are labeled
as the global best ($\textbf{x}_{\rm{g}}$). At $t>0$, the local best for each particle
is the set of coordinates that particle visited since $t=0$ that had the lowest associated
response value. Similarly, global best is the set of coordinates that resulted in the lowest
response value for any particle in the swarm since $t=0$.

\noindent 3) Compute particle velocities for a time step $\Delta t$ using
\begin{align}
    v^{i} (t+\Delta t) = C_1 R \left(x^{i}_\text{l} - x^{i}(t) \right) / \Delta t + 
    C_2 R \left(x^{i}_\text{g} - x^{i}(t)\right) / \Delta t.
\end{align}
Here,
 
$C_1$ and $C_2$ are the `self confidence' and `swarm confidence' factors, which
usually take a value between 0 and 2. Setting $C_2 = 0$ allows the particles to
search the space completely independently of each other (`exploration'), whereas
setting $C_1 = 0$ constrains the particles to move directly towards the global best
particle in the swarm (`exploitation'). In practice, setting both $C_1, C_2 > 0$
provides the best compromise between exploration and exploitation. The adjustable parameter
$R$ provides a random perturbation in the velocity to assist in HP space exploration.
$\Delta t$ can be adjusted to either speed up or slow down the particle.

\noindent 4) Update the new position of the particles using
\begin{align}
x^{i} (t+\Delta t) = x^{i}(t) + v^{i} (t+\Delta t),
\end{align}
and go to step 2 for the next iteration. We have developed our own PSO implementation
because of the need to include custom-built objective functions to target individual
stages in ML-PES training. Moreover, our problem requires a unique parallelization
strategy where both PSO algorithm and the objective function could run in parallel
in a hybrid fashion, to get the high-throughput required in HPO.

In this work, we use the QUIP/GAP~\cite{quip,bartok2015g} code built with the
TurboGAP~\cite{caro2019optimizing,turbogap} libraries, to train GAP models based on
the {\soapt} descriptor. We also use TurboGAP on its own to compute {\soapt}
descriptors.

%%%%%%%%%%%%%%%%%%%%%%%%%%%%%%%%%%%%%%%%%%%%%%%%%%%%%%%%%%%%%%%%%%%%%%%%%%%%%%%%%%%%%%%

\section{HPO Strategies}
\subsection{Optimization Cost}
HP optimization will be performed in several PSO iterations. In each iteration, a set
number of response variable evaluations will be made on the HP coordinates given
by the PSO algorithm. There are two strategic ways to optimize these HPs: 1) optimize
\{HP\}$^\text{SOAP}$ and \{HP\}$^\text{GAP}$ together by minimizing the RMSE of
the test set geometries; 2) optimize \{HP\}$^\text{SOAP}$ and \{HP\}$^\text{GAP}$
separately in two subsequent stages. Option 1 is the most obvious and straightforward
way to optimize the HPs, which by default takes into consideration the interaction
between \{HP\}$^\text{SOAP}$ and \{HP\}$^\text{GAP}$. By contrast, option 2 is
based on the assumption that the interaction between the two sets of HPs is
negligible so that \{HP\}$^\text{SOAP}$ can be optimized first, followed by the
optimization of \{HP\}$^{\rm{GAP}}$. Option 2 gives us a distinct advantage in terms
of computational effort needed to optimize the HPs. The number of \{HP\}$^\text{SOAP}$
is usually larger than the number of \{HP\}$^\text{GAP}$, whereas the time required
to train a GAP is much larger than the time needed to evaluate SOAP descriptors. If
we could find a way to choose the best \{HP\}$^\text{SOAP}$ to describe the data set
cheaply, i.e., without training GAPs, we would be able to search the relatively smaller
GAP HP space faster and more thoroughly. Once the optimal SOAP basis set is chosen,
techniques such as those mentioned in Ref.~\cite{imbalzano2018automatic} can be
applied to further reduce the number of descriptor functions within the basis set without sacrificing the quality of the descriptors.

For GAP, the RMSE as quality parameter is an obvious objective function to minimize.
However, an appropriate objective function for finding the best SOAP is not obvious.
This means that we need to devise a suitable objective function to optimize SOAP for
a given database. Since the prospect of training-free HPO is very enticing, we will
evaluate both options in this manuscript.

In strategy 1, the response variable is the RMSE ($\Gamma_E$, $\Gamma_F$) of the
test set and we will use the full HP search space (16 dimensional) to find the
minima.

In strategy 2, optimization is done in two stages. In the first stage, SOAP HPs
are optimized in the SOAP HP sub-space (12 dimensional, including $\zeta$) by
considering a SOAP kernel distribution property as the response variable (e.g.,
maximizing its standard deviation).

In the second stage of strategy 2, GAP HPs are optimized in the GAP HP
sub-space (4 dimensional) with fixed SOAP HPs obtained from the first stage and
using the RMSE of the test set as the response variable. 

Strategy 2 requires relatively high computational effort per iteration (stage 1
and 2 combined) when compared to strategy 1. However, stage 1 of strategy 2
requires less computational effort per iteration than strategy 1 since SOAP
evaluation is much cheaper than GAP training. Also, stage 2 of strategy 2 will
require a lower number of iterations to converge than strategy 1 because the HP
search space is much smaller for the former. This means that we will need to train
less GAPs using strategy 2 than using strategy 1 in the entire PSO run. Therefore,
there are advantages in using strategy 2 if GAP training is expensive. On the other
hand, if the GAP training and RMSE computation are cheap, then strategy 1 should
be optimal. Note that, in both strategies, 
$\sigma_F$ need not be optimized if forces are not included in the training. 

\subsection{C Dimer System}
To test our HPO strategies, as a first example, we choose a `toy' data set of
an isolated C dimer. Note that the scalar distance between the two C atoms is
enough to completely describe the C dimer system. However, we use the SOAP
descriptor in this analysis simply as a proof of principle and to compare the
results with more complex data sets in the next section. There are 30 structures
in our dimer data set, with C-C distances ranging from 0.8~\AA{} to 3.7~\AA{},
increasing in steps of 0.1~\AA{}. Because of symmetry, both C atoms in a given
dimer structure will have the same SOAP. The optimal SOAP will change
systematically as we go from contracted dimer to stretched (dissociated) dimer,
and this difference should be picked up by the SOAP kernel, i.e., the dot product
between SOAP vectors of C atoms from any two structures raised to the
power of $\zeta$. While $\zeta$ is not formally a SOAP HP, we include it in
\{HP\}$^\text{SOAP}$ since the kernel distribution is dependent on it.

\subsubsection{Importance of \{HP\}$^\text{SOAP}_\text{S}$}
First of all, we want to showcase the impact of the shape affecting SOAP HPs
on SOAP kernels and why it is important to optimize them for the data set at hand.
Let us compare two sets of \{HP\}$^\text{SOAP}$s with fixed values for $l_\text{max} = 6$,
$n_\text{max} = 6$ and rc$_\text{h} = 5$~\AA{}, keeping $N^\text{SOAP}$ constant at
147, and different values for the other HPs as listed in Table~\ref{tab:HPdimer}. 
 \begin{table}[ht]
 \small
     \centering
     \begin{tabular}{l c c}
     \hline
        \{HP\}$^{\rm{SOAP}}$  &  Set 1 & Set 2\\
        \hline
        n$_{\rm{max}}$  & \multicolumn{2}{c}{6}\\
        l$_{\rm{max}}$ & \multicolumn{2}{c}{6}\\
        rc$_{\rm{h}}$   & \multicolumn{2}{c}{5.0}\\
        rc$_{\rm{s}}$     & 4.21 & 3.40 \\
        $\sigma_r$      & 0.69 & 0.13\\
        $\sigma_t$      & 0.74 & 1.39\\
        a$_r$            & 0.88 & 0.13\\
        a$_t$            &0.69 & 0.51\\
        a              &0.64 & 1.45\\
        cw              &0.53 & 1.26\\
        re              &0 & 2 \\
        \hline
        $\zeta$ &1&1\\
        Mean & 0.994 & 0.358\\
        Std & 0.008 & 0.379\\
        \hline
        $\zeta$ &10&10\\
        Mean & 0.944 & 0.132\\
        Std & 0.075& 0.280\\
        \hline
     \end{tabular}
     \caption{A list of two distinct sets of SOAP HPs with fixed values for \{HP\}$^{\rm{SOAP}}_{\rm{P}}$ along with the mean and standard deviation of the corresponding SOAP kernels of C dimer data set.}
     \label{tab:HPdimer}
 \end{table}
The table also lists the resulting mean and standard deviation (std) of the
corresponding kernel distribution at $\zeta = 1$ (linear kernel in the SOAP
components) and at $\zeta = 10$. Note that these two sets of HPs are somewhat
arbitrary, having been chosen for the sole purpose of argumentation of how different,
naively chosen, descriptor HP sets can lead to drastically different kernel
distributions, and therefore drastically different ML model performance.

\begin{figure*}
    \centering
    \includegraphics[scale=0.4]{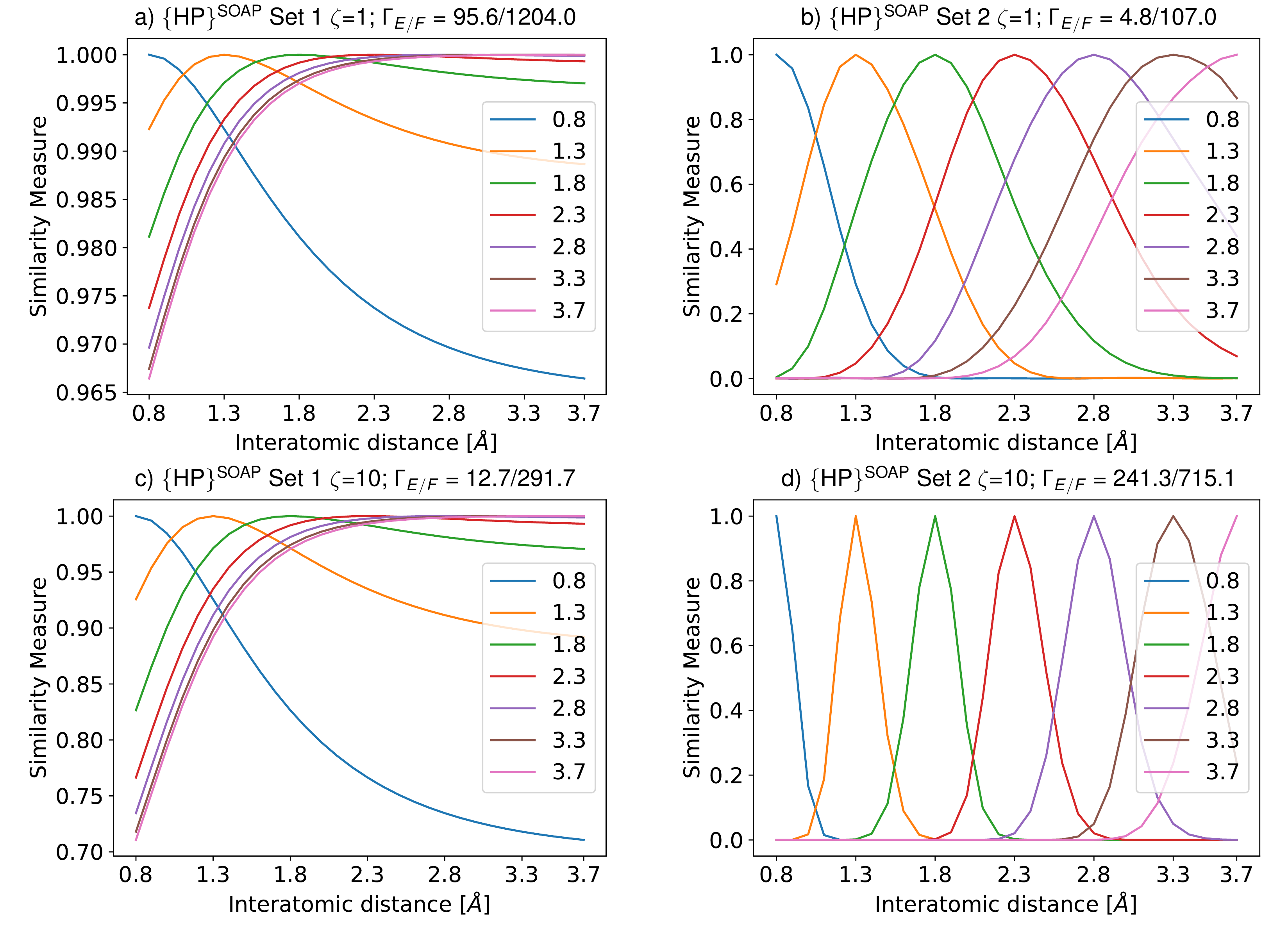}\\
    \caption{Similarity measure of the C environments in 7 different structures
    (with C-C distance in \AA{} given in the legend) compared to all the unique
    environments in the data set (all interatomic distances in $x$-axis) for the
    two sets of \{HP\}$^\text{SOAP}$ are shown in panels (a) for set 1 and (b)
    for set 2 with $\zeta=1$. Similar curves for $\zeta = 10$ are given in panels
    (c) for set 1 and (d) for set 2. $\Gamma_E$ is given in meV/atom and
    $\Gamma_F$ is given in meV/\AA. Note the different $y$-axis ranges in the plots.}
    \label{fig:c-cenv}
\end{figure*}

For the two SOAP HP sets, the similarity curves of 7 selected
structures, as distinct from each other as possible in the data set, are compared
in Figure~\ref{fig:c-cenv} for $\zeta=1$ and $\zeta=10$. 
 \begin{figure*}
    \centering
    \includegraphics[scale=0.4]{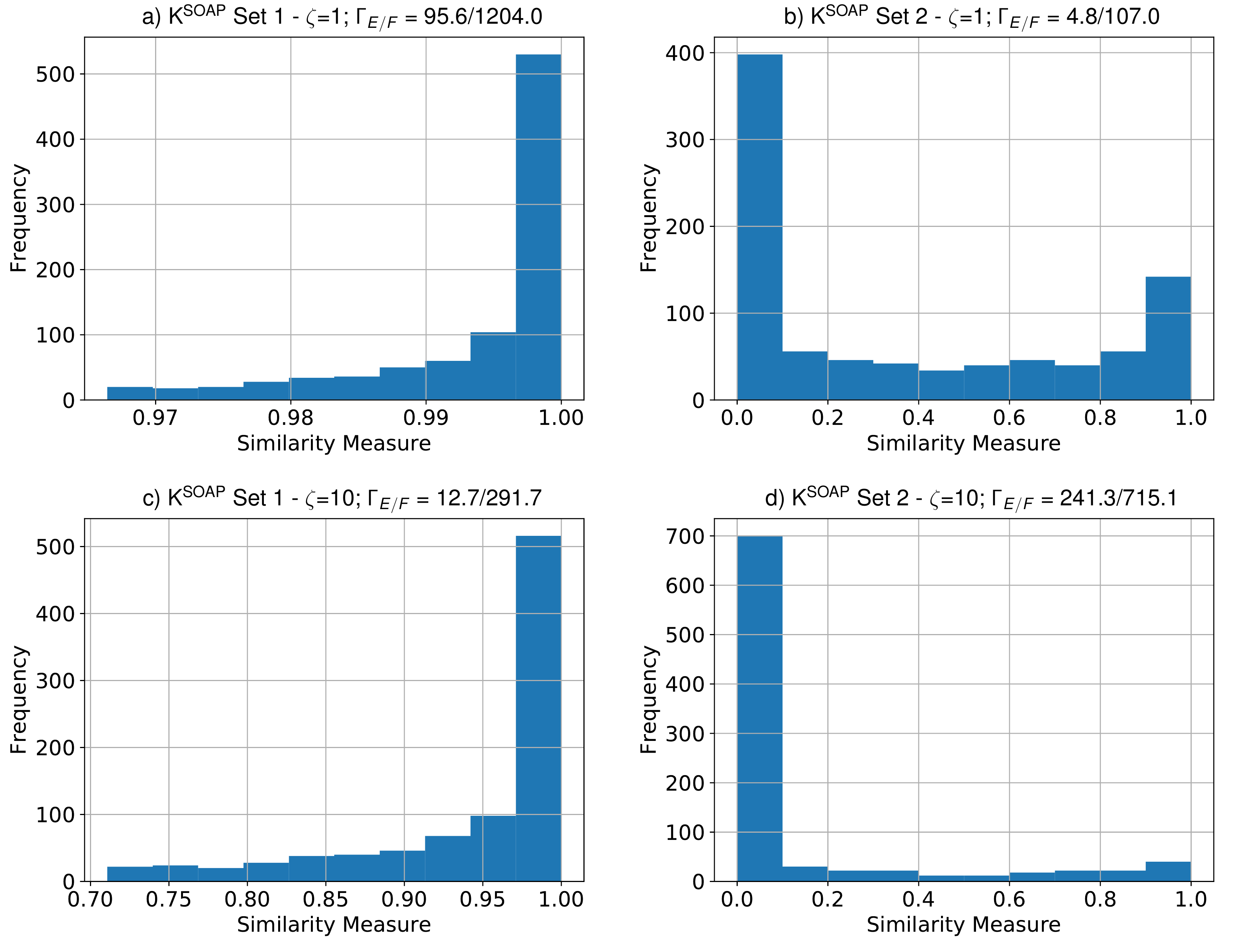}\\
    \caption{Kernel distributions given by the two sets of \{HP\}$^\text{SOAP}$ are
    shown in panels (a) for set 1 and (b) for set 2 with $\zeta=1$. Similar distributions
    for $\zeta = 10$ are given in panels (c) for set 1 and (d) for set 2. $\Gamma_E$ is
    given in meV/atom and $\Gamma_F$ is given in meV/\AA.}
    \label{fig:c-cenv2}
\end{figure*}
The corresponding kernel distributions are given in Figure~\ref{fig:c-cenv2}. Please note that we use automatic range for x-axis based on the data as it also helps to see the range of similarity measures the kernel produces. As
mentioned earlier, the kernel gives a similarity measure between 1 and 0, with 1
being exactly the same environment (up to symmetry operations) and 0 being nothing
alike. Essentially, $\zeta$ has the overall effect of scaling the kernel value.
Larger values of $\zeta$ emphasize the differences between atomic environments (make
the kernel `sharper'). From Table~\ref{tab:HPdimer}, Figures~\ref{fig:c-cenv}a
and~\ref{fig:c-cenv2}a, we see that set 1 has a very narrow kernel distribution at
$\zeta = 1$ ($\text{std} = 0.006$), and the mean close to 1 ($\text{skew} = 1.6$)
means that the SOAP vectors from all the structures are very similar. A distribution
that is symmetric about its mean value will have $\text{skew} = 0$. Increasing the value
of $\zeta$ broadens the kernel distribution of set 1 ($\text{std} =  0.06$,
$\text{skew} = 1.5$) by scaling down the kernel values, as seen in Table~\ref{tab:HPdimer},
Figures~\ref{fig:c-cenv}c and~\ref{fig:c-cenv2}c. The numbers given above suggest that
set 1 HPs do not result in a highly resolved SOAP since (i) all kernel values are
$> 0.90$ for the linear kernel (Figure~\ref{fig:c-cenv2}a) and $> 0.70$ for $\zeta = 10$
(Figure~\ref{fig:c-cenv2}c); (ii) environments with C-C distance between 2.8~\AA{} and
3.7~\AA{} look very similar and the broadened kernel distribution at $\zeta = 10$ did
not change the shape of the similarity curves either (Figures~\ref{fig:c-cenv}a
and~\ref{fig:c-cenv}c).

In contrast with set 1, set 2 shows a reasonably broad kernel distribution ($\text{std} =
0.38$) already at $\zeta = 1$, with the kernel values spanning the full range
between 0 and 1 ($\text{skew} = 0.5$), as seen in Figures~\ref{fig:c-cenv}b
and~\ref{fig:c-cenv2}b. Here, set 2 linear kernel provides clearly distinguishable
similarity curves for each environment.

Further increasing $\zeta$ in set 2 leads to continued down scaling of the kernel
values ($\text{std} = 0.28$; $\text{skew} = 2.08$) and results in the similarity
measures falling steeply for almost all environments (Figures~\ref{fig:c-cenv}d
and~\ref{fig:c-cenv2}d). Set 2 HPs with large $\zeta$ values will therefore result
in under-fitting when used to train a GAP. For quantitative understanding, we trained
C dimer GAPs on DFT energies and forces using the above four SOAP HP sets (with
$\delta = 1.0$~eV, $\sigma_E$= 2 meV/atom and $\sigma_F$=20 meV/\AA). The corresponding
energy and force RMSEs are shown in the respective panels in Figures~\ref{fig:c-cenv}
and~\ref{fig:c-cenv2}. We see a direct correlation between the changes in the std of
the kernel distribution and the training errors: large std leads to low errors and
vice-versa. 

\begin{figure*}
    \centering
    \includegraphics[scale=0.43]{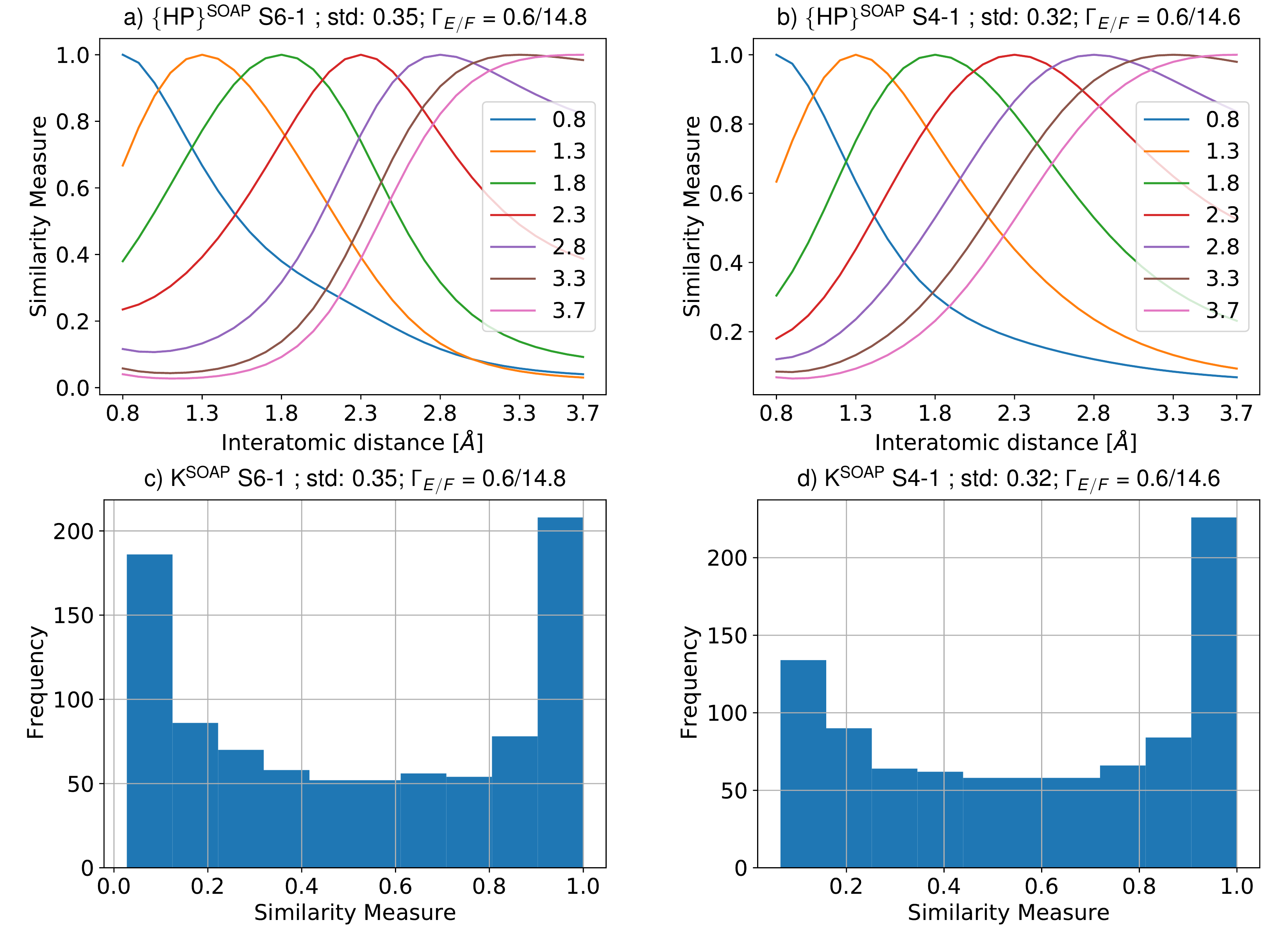}\\
    \caption{Similarity curves of 7 C environments given by the HP sets S6-1 and S4-1 are given in panels a) and b), respectively. The corresponding kernel distributions are shown in panels c) and d). $\Gamma_E$ is given in meV/atom and $\Gamma_F$ is given in meV/\AA.}
    \label{fig:c-cenv64}
\end{figure*}

Based on the above observations, std of SOAP kernel distribution is chosen as a
response variable in optimizing the shape-affecting SOAP HPs. The distribution in
Figure~\ref{fig:c-cenv2}b, although spanning the entire kernel range, is skewed
towards low similarity values. We will find out in the next section whether std
is a good universal response variable for SOAP optimization or not. More importantly,
this tells us that choosing a large basis set and cutoff (\{HP\}$^\text{SOAP}_\text{P}$)
does not automatically guarantee a reasonable SOAP description of the system without
the careful selection of \{HP\}$^\text{SOAP}_\text{S}$. Choosing the best
\{HP\}$^\text{SOAP}_\text{S}$ is non-trivial even for the simple C dimer data set,
as we have discussed above, and is best done by efficient optimization algorithms
such as the parallel PSO described in the previous section. We will look at more
complex data sets in later sections.

\subsubsection{Strategy 1}
We begin our HP optimization studies of C dimer data set with strategy 1,
in which both SOAP and GAP HPs are optimized at the same time by minimizing a linear
combination of the energy and force errors, here chosen as
$\Gamma_E$+($\Gamma_F$/30). This way the code optimizes HPs by minimizing both
energy and force errors at the same time. Since the magnitude of the energy error is
usually smaller than the magnitude of the force error (when expressed in eV and eV/{\AA},
respectively), we downscale the force error contribution by a factor of 30, to give
roughly equal importance to both errors in the response variable. We initialized 16
particles in the swarm and selected a self-confidence to swarm-confidence ratio of 1:2
to put more emphasis on exploitation of the global best HP coordinates in each swarm
step. We would also like to see if a smaller SOAP basis set could offer similar quality
of GAPs as the larger ones.

Therefore, we started two PSO runs, one with fixed values of $l_\text{max} =
n_\text{max}=6$ (S6-1) and the other with $l_\text{max} = n_\text{max} = 4$ (S4-1),
to keep the total number of SOAP components $N^\text{SOAP}$ constant at 147 and 50,
respectively. We also fixed the hard cutoff at 5~\AA{} and randomly initialized
the values of the other HPs within the bounds given in Table~\ref{tab:HPOdimerf}.

\begin{table}
\small
     \centering
     \begin{tabular}{l c c c c c c}
     \hline
        HP$^{\rm{SOAP}}$  &  LB & UB & S6-1 & S4-1 & S6-2& S4-2\\
        \hline
        n$_{\rm{max}}$  & \multicolumn{2}{c}{6/4} &6 &4 &6 &4 \\
        l$_{\rm{max}}$ & \multicolumn{2}{c}{6/4} &6 &4&6 &4 \\
        rc$_{\rm{h}}$   & \multicolumn{6}{c}{5.0}\\
        rc$_{\rm{s}}$    & 4.0 & 5.0 &4.60& 4.61&4.49& 4.87\\
        $\sigma_r$       & 0.1 & 2.0 & 0.38& 0.24& 0.23& 0.15\\
        $\sigma_t$       & 0.1 & 2.0 & 1.28& 1.43& 1.22& 1.08\\
        a$_r$            & 0.0 & 2.0 & 1.06& 0.59& 1.20& 0.56\\
        a$_t$            & 0.0 & 2.0 & 0.05& 1.20& 0.32 & 0.18\\
        a                & 0.0 & 2.0 &1.41& 0.98&1.52& 0.34\\
        cw               & 0.0 & 2.0 &0.81& 0.92&1.26& 0.73\\
        re               & 0 & 2 & 1&1& 1&2\\
        $\zeta$          &2 &10 &5.52&4.60&7.21&3.79\\
        $\delta$         & 0.1 &5.0 &3.50&2.49&4.24&4.31\\
        $\sigma_E$       &  0.001&0.100&0.056&0.061&0.001&0.001\\
        $\sigma_F$       & 0.001&0.100&0.001&0.001&0.014&0.024 \\\\
        K mean           & & & 0.51&0.58& 0.51& 0.49\\
        K std            & & & 0.35&0.32&0.37 & 0.39\\
        $\Gamma_E$ & \multicolumn{2}{c}{[meV/atom]}  &0.6& 0.6&0.6& 0.5\\
        $\Gamma_F$ & \multicolumn{2}{c}{[meV/\AA]}   &14.8&14.6&15.3&15.1\\
        \hline
     \end{tabular}
     \caption{A list of upper and lower bounds (UB and LB) of HPs used in the PSO run along with the best full HP sets for C dimer obtained from both strategy 1 (S6-1 and S4-1) and strategy 2 (S6-2 and S4-2). K mean and K std values correspond to the mean and standard deviation of the respective SOAP kernel distributions.}
     \label{tab:HPOdimerf}
 \end{table}

The table also includes the optimized HPs for the two sets along with the minimum
response variable (RMSEs) and mean and std of corresponding SOAP kernels.

We find that, for this extremely simple system, the two differently sized SOAP
can be optimized to provide GAPs with
almost the same accuracy, energy RMSE of 0.6 meV/atom and force RMSE of
$14.7 \pm 0.1$ meV/\AA. The SOAP kernel distribution and similarity curves of
the 7 distinct C environments are shown in Figure~\ref{fig:c-cenv64} for the
optimized SOAP HPs.

From Figures~\ref{fig:c-cenv64}a and~\ref{fig:c-cenv64}b, we find that both HP
sets differentiate the C environments very well. From the kernel distributions in
Figures~\ref{fig:c-cenv64}c and~\ref{fig:c-cenv64}d, we find that the similarity
measures span the entire kernel range and the distribution is also broad (std of 0.32
for S4-1 and 0.35 for S6-1) and more symmetric than those in Figure~\ref{fig:c-cenv2}.

\subsubsection{Strategy 2}
We proceed with the first stage of strategy 2 in which we employed a PSO run with 16
particles to obtain optimal \{HP\}$^\text{SOAP}_\text{S}$ values for this simple C dimer
data set by maximizing the standard deviation of the SOAP kernel distribution. Again,
we test the two differently sized SOAP bases with $l_\text{max} = n_\text{max} = 4, 6$
(S4-2 and S6-2, respectively) at fixed hard cutoff of 5~\AA{} and random initialization
of the values of the other HPs within the bounds given in Table~\ref{tab:HPOdimerf}. 

\begin{figure}
    \centering
    \includegraphics[scale=0.45]{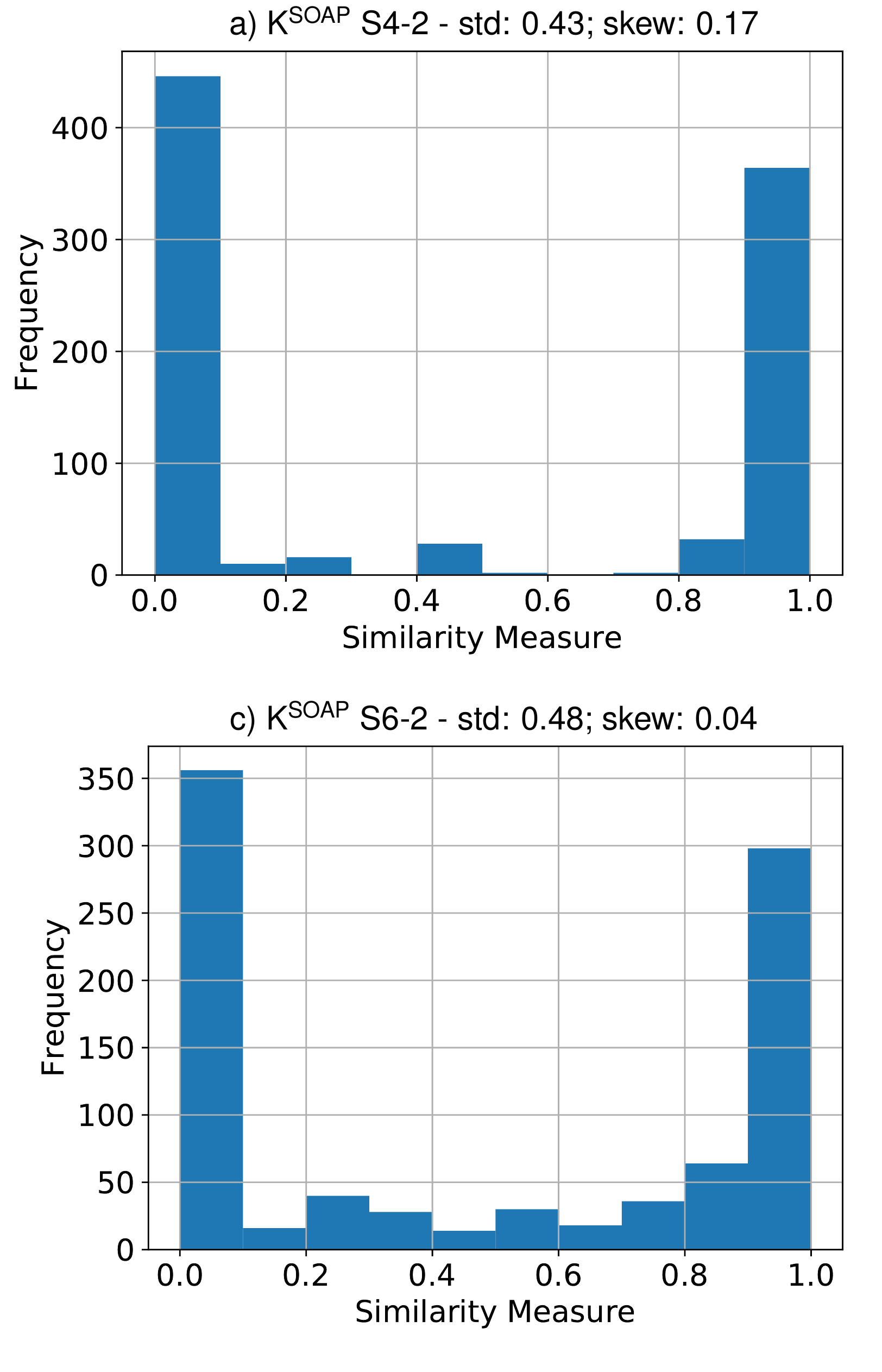}
    \caption{The SOAP kernel distribution of S4-2 (a) and S6-2 (b) obtained from HPs that maximizes the standard deviation of the kernel distributions.  }
    \label{fig:stdopt}
\end{figure}

  \begin{figure*}
    \centering
    \includegraphics[scale=0.45]{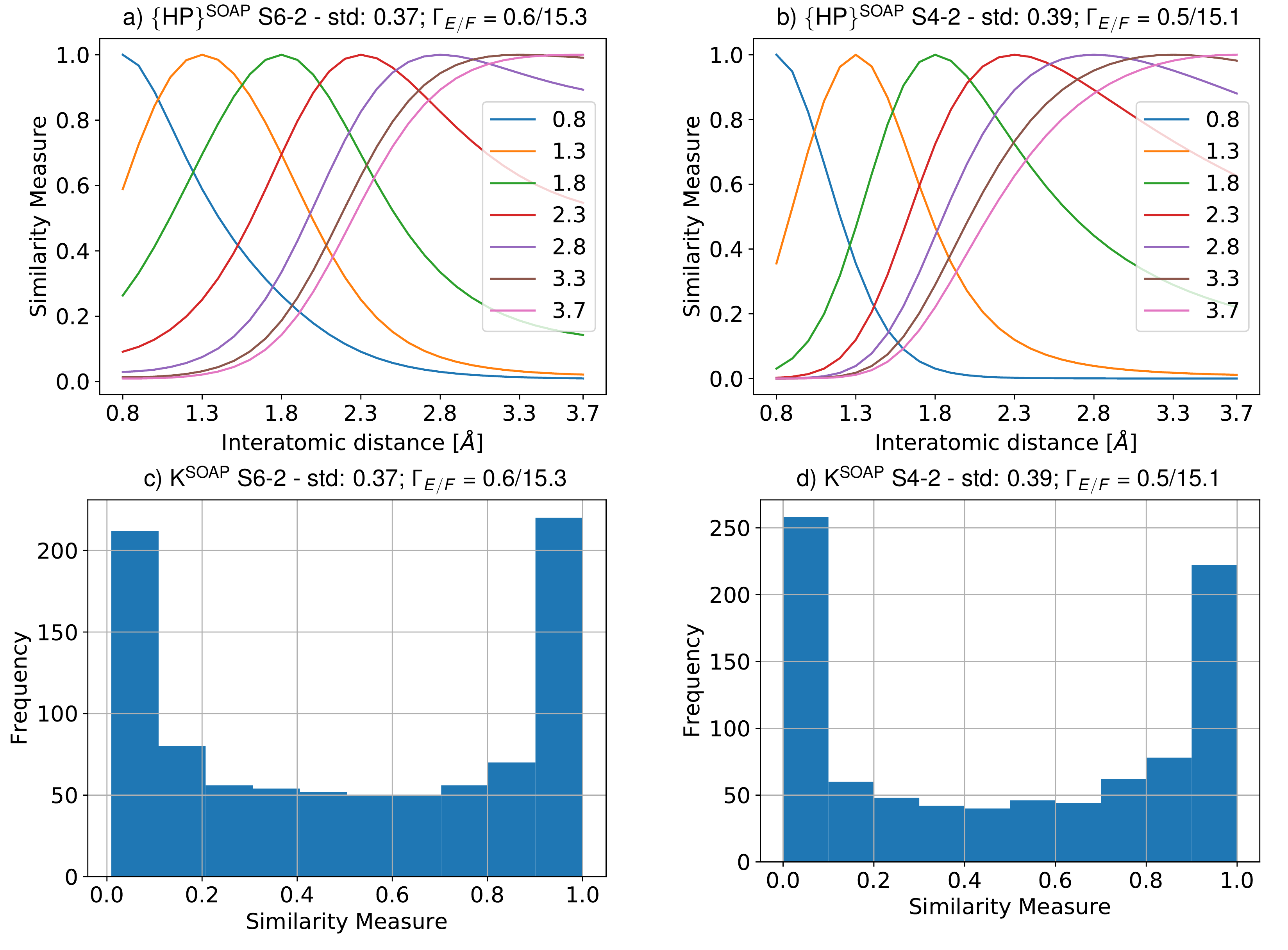}
    \caption{Similarity curves of selected 7 C environments for the SOAP HP sets S6-2 and S4-2 are shown in panels (a) and (b). The corresponding kernel distributions are given in panels (c) and (d). $\Gamma_E$ is given in meV/atom and $\Gamma_F$ is given in meV/\AA.}
    \label{fig:ccswarm}
\end{figure*}

Maximizing the std of the kernel distribution ultimately led to increased tail
weights, i.e., the contributions close to 0 and 1 increased while the
contributions in between diminished, as shown in Figure~\ref{fig:stdopt}. Using such
kernels in GAP training will result in under fitting. This implies that while a broad
kernel distribution is important to better distinguish between chemical environments,
it is not necessarily true that the broadest kernel distribution will give the best
possible GAP. Therefore, instead of maximizing only the std of the kernel distribution,
we chose to optimize the HPs by simultaneously maximizing std and minimizing the skew
of the distribution.

The similarity curves of the 7 C-dimer environments and the kernel distributions
(similar to Figure~\ref{fig:c-cenv64}), obtained from the SOAP HPs that are optimized
using the revised response variable, are shown in Figure~\ref{fig:ccswarm}; the
corresponding HP coordinates are given in Table~\ref{tab:HPOdimerf}. From the table,
we find that the kernel distributions of both sets look very similar in terms of the
mean and std values, despite the fact that S4-2 has roughly 100 less SOAP components
than S6-2. This suggests that HPO can be used to find the optimal size and shape of
SOAP basis set \textit{for a given data set}. In the next section we will find out
whether this strategy also works for complex data sets.

To put our predictions to the test, we applied the PSO code to optimise GAP HPs
(stage 2) within the ranges given in Table~\ref{tab:HPOdimerf}, while keeping the
SOAP HPs fixed for the two cases. We used 20 particles per swarm, the same confidence
ratio of 1:2 as before, and set the target response variable to $\Gamma_E$+($\Gamma_F$/30).
We used both energies and forces to train the GAPs. From Table~\ref{tab:HPOdimerf}, we
see that both HP sets, S4-2 and S6-2, result in energy errors below 1 meV/atom and
force errors below 20 meV/\AA. The quality of the fits obtained from strategy 2 is
comparable to those obtained from strategy 1.
This indicates that our divide-and-conquer approach (strategy 2), to optimize SOAP and GAP
HPs separately, works for this toy C dimer system.

\cleardoublepage
\subsection{Complex Data Sets}
So far, we have tested our two proposed HPO strategies on a toy dimer system. We have
identified that kernel distribution properties such as skewness and std can be used
as response variables in obtaining optimal SOAP for a given data set. Now, we apply
the two strategies to optimize SOAP and GAP HPs for already published amorphous
carbon~\cite{deringer2017machine} and $\alpha$-iron~\cite{Feds} data sets. It is to
be noted that the optimal kernel distribution for condensed-phase systems may not
look symmetric as in the case of the C dimer data set, as there will be a lot more
environments that look similar (kernel value close to 1) than the ones that are completely
dissimilar (kernel value close to 0). It will be interesting to see what kind of kernel
distributions we arrive at by maximizing its std and minimizing its skewness.

The amorphous carbon (a-C) database consists of 4080 structures out of which there
are 3070 bulk amorphous carbon structures, 356 crystalline structures, 624 amorphous
surface geometries and 30 C dimer geometries (those studied in the previous section).
There are a total of 256,628 C atomic environments in this database, 
and it includes several high energy structures with very high forces (ca. 100~eV/\AA).
Therefore, we should expect relatively larger energy and force errors for this system as
compared to the simple dimer data set. The $\alpha$-Fe database consists of 12,171
structures in total, with 152,293 Fe atomic environments. The largest force value in
this database is ca. 30~eV/\AA.

If we denote the total number of atomic environments in a database be $N_\text{env}$,
then the total number of unique kernels that can be constructed is $N_\text{env}
(N_\text{env}-1)$ (where we have removed the trivial self-similarities). In our databases,
this leads to a total of similarity checks in the order of $>10^{10}$. Handling this number
of data points is intractable computationally (both in terms of CPU time and memory). Therefore, we
have used 3000 randomly selected environments to construct the SOAP kernels, for a total
of $3000 N_\text{env}$ similarity checks, which is much more manageable. For the GAP fits,
4030 (a-C) and 4500 ($\alpha$-Fe) sparse environments ($n_\text{sparse}$) are chosen
by $CUR$ decomposition,~\cite{mahoney2009cur} a low-rank matrix decomposition method,
with 20\% (a-C) and 30\% ($\alpha$-Fe) of the structures in the data set randomly chosen
per configuration type to make up the test set (TE), the rest making up the training
set (TR). We have simply used a subset of the full data set as the
test set, but a comprehensive test suite tailor-made for the specific system can also
be employed here. Moreover, we reiterate that the GAP fits are trained only on reference
energies, for computational efficiency, but the response is taken from both energy
and force errors.

We tested the four optimal HP sets obtained in the previous section to find out whether
they would perform well also for the condensed-phase system. We found that all of the
four sets led to large errors for the a-C data set. Thus, there is no guarantee that
the SOAP HPs that work well for one data set will be optimal for a different data set.
Therefore, we must, in principle, optimize SOAP and GAP HPs for every new data set.

\subsection{Amorphous Carbon Database}
\subsubsection{Strategy 1}
For the complex a-C database, we test strategy 1 in which we optimized both SOAP
and GAP HPs together within the bounds given in Table~\ref{tab:HPOaC}. Note
that the GAPs are trained only on reference energies and not forces. Since we do not
train with force information, $\sigma_F$ HP need not be optimized.
 \begin{table}[h]
 \small
     \centering
     \begin{tabular}{l c c c c c c}
     \hline
        HP$^{\rm{SOAP}}$  &  LB & UB & aC6-1 & aC4-1 & Fe6-1 & Fe4-1\\
        \hline
        n$_{\rm{max}}$   & \multicolumn{2}{c}{6/4} &6 &4 &6 &4 \\
        l$_{\rm{max}}$   & \multicolumn{2}{c}{6/4} &6 &4 &6 &4 \\
        rc$_{\rm{h}}$    & \multicolumn{6}{c}{5.0}\\
        rc$_{\rm{s}}$    & 4.0 & 5.0 & 4.77 & 4.42 & 4.26 & 4.27 \\
        $\sigma_r$       & 0.1 & 2.0 & 0.63 & 0.48 & 0.38 & 0.35 \\
        $\sigma_t$       & 0.1 & 2.0 & 0.50 & 0.17 & 0.68 & 0.22 \\
        a$_r$            & 0.0 & 2.0 & 0.003 & 0.03 & 0.05 & 0.81 \\
        a$_t$            & 0.0 & 2.0 & 0.02 & 0.12 & 0.20 & 0.12 \\
        a                & 0.0 & 2.0 & 1.30 & 1.98 & 1.20 & 1.36 \\
        cw               & 0.0 & 2.0 & 0.46 & 0.90 & 1.23 & 0.51 \\
        re               & 0   & 2   & 0    & 2    & 1 & 2 \\
        $\zeta$          & 2   & 10  & 7.30 & 5.39 & 6.18 & 8.20 \\\\
        K mean           &     &     & 0.87 & 0.86 & 0.99 & 0.99 \\
        K std            &     &     & 0.12 & 0.12 & 0.04 & 0.04 \\
        K skew           &     &     & 1.63 & 1.56 & 8.31 & 11.2 \\\\
        n$_{\rm{sparse}}$   & \multicolumn{2}{c}{} &\multicolumn{2}{c}{4030} &\multicolumn{2}{c}{4500} \\
        $\delta$         & 0.1 & 5.0 & 2.72 & 2.54 & 4.95 & 3.84 \\
        $\sigma_E$       & 0.001 & 0.100 & 0.010 & 0.015 & 0.001 & 0.001 \\\\
        $\Gamma_E$ & \multicolumn{2}{c}{[eV/atom]}  &0.032  & 0.037 & 0.003 & 0.005 \\
        $\Gamma_F$ & \multicolumn{2}{c}{[eV/\AA]}   &1.211 & 1.172 & 0.093 & 0.124 \\
        \hline
     \end{tabular}
     \caption{A list of upper and lower bounds (UB and LB) of SOAP HPs and GAP HPs along with the kernel and GAP quality estimates for sets S4-1 and S6-1 representing amorphous carbon and $\alpha$-Fe databases.}
     \label{tab:HPOaC}
 \end{table}
Table~\ref{tab:HPOaC} also contains the optimized HPs for the two SOAP basis sizes,
aC4-1 and aC6-1 with $\text{rc}_\text{h}$ fixed at 5~\AA. We were able to find optimal
HPs for each of the sets providing similar test set errors. This suggests that by
optimizing the HPs, a small SOAP basis can perform similarly to a larger basis, even
for a complex data set. This amorphous carbon database consists of liquid carbon
geometries with some large forces, thus the large errors and it is consistent with
the values reported in the original paper.~\cite{deringer2017machine} In the original
paper, the final potential is a combination of SOAP, 2-body and 3-body contributions,
and $\text{rc}_\text{h}$ was set to 3.7~\AA. In our PSO test, we have only used SOAP
contributions. However, it is straightforward to include the 2-body and 3-body HPs
here, which will lead to a nominal increase in the computational cost per iteration. 

\begin{figure*}
    \centering
    \includegraphics[scale=0.45]{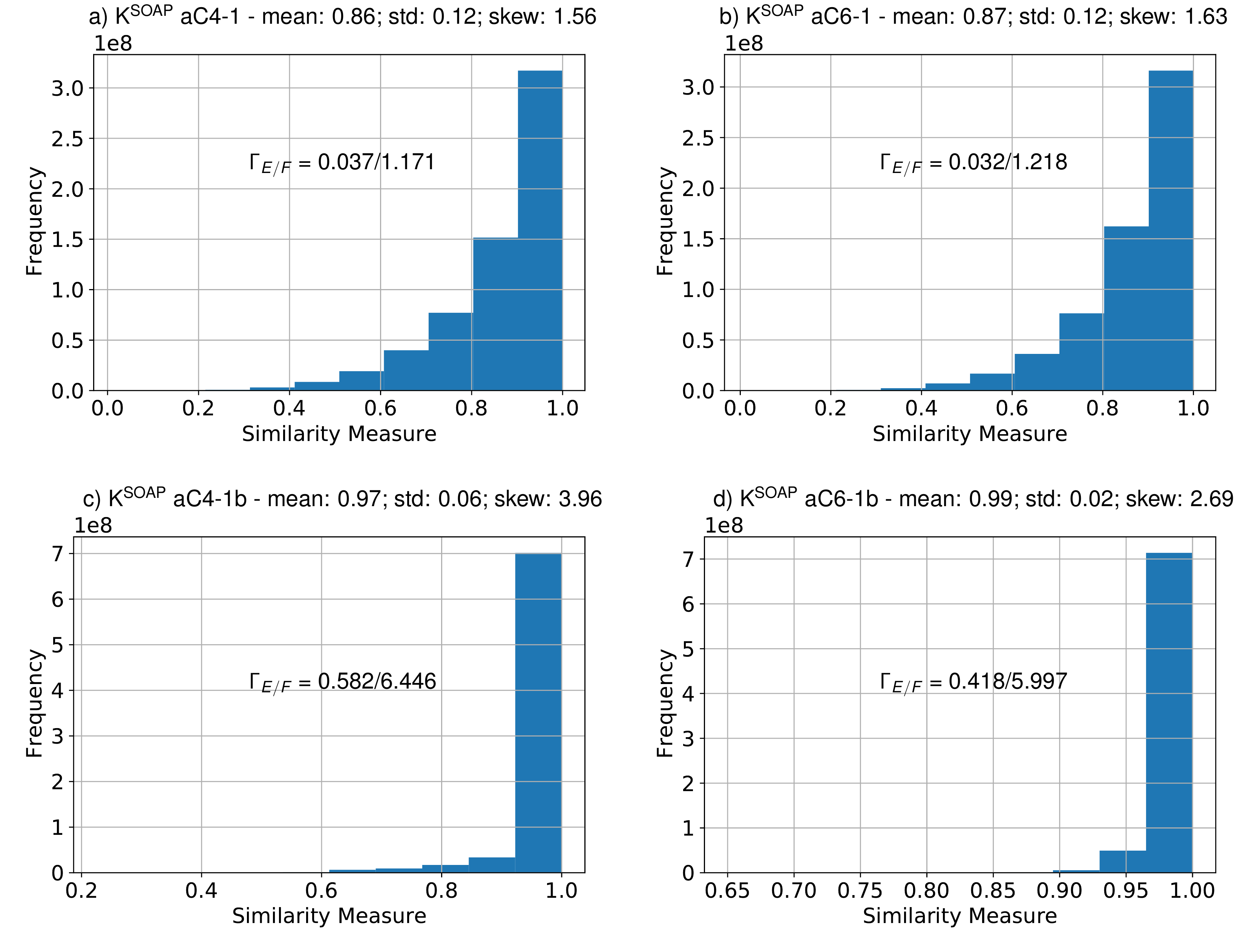}
    \caption{The SOAP kernel distribution of aC4-1 (a) and aC6-1 (b) obtained from HPs that minimized the test set RMSE. c) and d) are examples for non-optimal HPs that show a narrow kernel distribution.}
    \label{fig:Kac64}
\end{figure*}
\begin{figure*}
    \centering
    \includegraphics[scale=0.5]{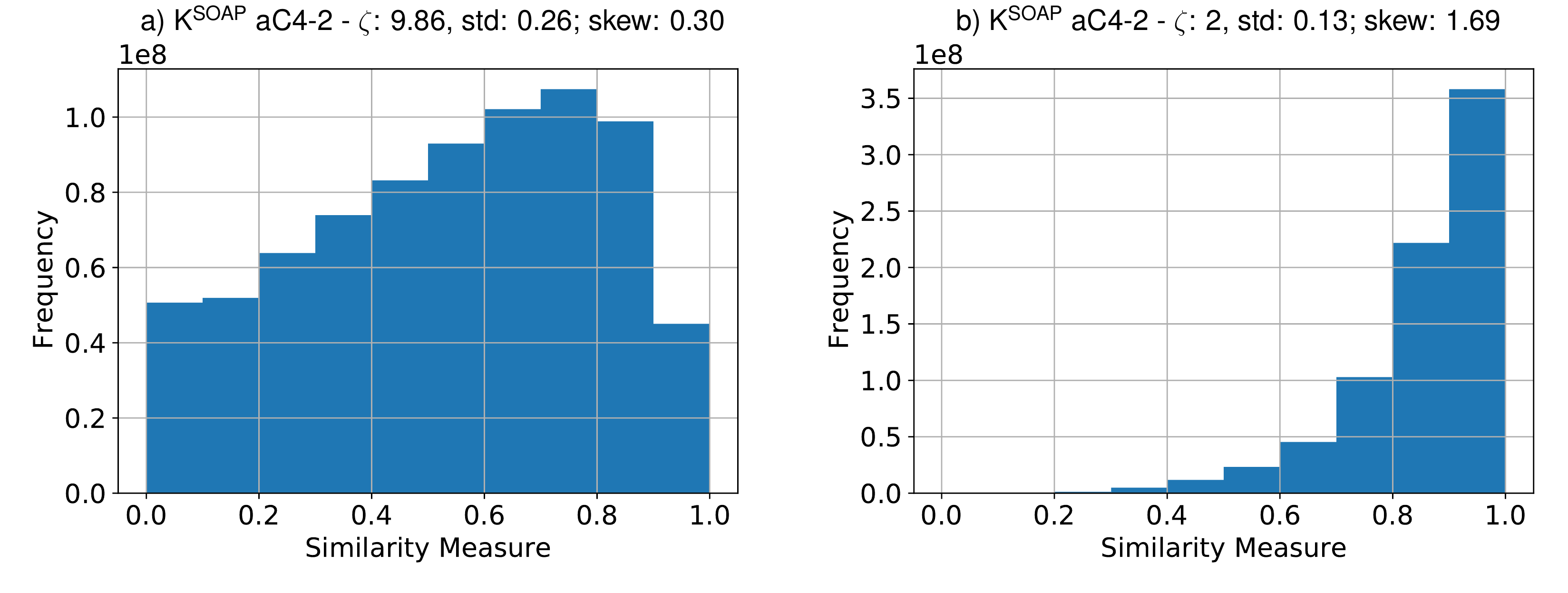}
    \caption{a) The SOAP kernel distribution obtained from SOAP HPs, aC4-2, that maximized the std and minimized the skewness of the distribution. Mean: 0.54, std: 0.26, skew: 0.30. SOAP HPs are $\sigma_r$=0.17, $\sigma_t$=0.50, a$_r$=0.21, a$_t$=0.03, a=1.75, cw=0.59, re=1, rc$_{\rm{s}}$=4.42 and $\zeta$=9.85. b) panel shows SOAP kernel distribution same as in panel a but with $\zeta$=2.}
    \label{fig:Kac4-2}
\end{figure*}

As the next step, we look at the SOAP kernel distribution given by the two sets
of optimal HPs in Figures~\ref{fig:Kac64}a and~\ref{fig:Kac64}b. As we anticipated,
the kernel distribution is weighted heavily towards 1 with a mean value ca. 0.87
for both sets. We also notice that the range of similarity measure extends all the way
down to 0 in both cases. Note that we use automatic range for x-axis based on the data.
This is due to the fact that the frequency for large similarity measures are very
high compared to that of small similarity measures. If 0.0 is visible in the x-axis
then there is at least one kernel with 0 similarity.

To validate the use of std and skew of the kernel distribution as the response
variable in stage 1 of strategy 2, we compare the kernel distributions of the HP
configurations that gave large errors in Figures~\ref{fig:Kac64}c
and~\ref{fig:Kac64}d. We find the non-optimal HPs (aC4-1b and aC6-1b) gave
narrow kernel distributions, similar to the C dimer case, along with large skew,
and they did not span the entire kernel space either. It appears that our SOAP
optimization objectives, maximizing std and minimizing skewness of the kernel
distribution, might work for the condensed phase systems as well.   

\subsubsection{Strategy 2}
We execute stage 1 of strategy 2 using the smaller SOAP basis, aC4-2, as the test case
to optimize SOAP HPs for the a-C data set within the bounds given in
Table~\ref{tab:HPOaC}. By maximizing the std of the kernel distribution to 0.26 and
minimizing its skewness to 0.30 using PSO, we obtained a set of optimal SOAP
HPs, aC4-2, given in Figure~\ref{fig:Kac4-2}a (HP values in figure caption).

The value of $\zeta$ is pushed towards the upper bound of its chosen range by the
optimization procedure, which is not surprising as increasing $\zeta$ down scales
the kernel values and widens their distribution. For comparison, we recomputed the
kernel distribution by setting the $\zeta$ value to 2 while keeping the other HPs
constant (in Figure~\ref{fig:Kac4-2}b). This distribution has a mean of 0.85,
std of 0.13 and skew of 1.68, which is similar to the optimal distribution found in
strategy 2 (Figure~\ref{fig:Kac64}a).

Fixing the SOAP HPs, aC4-2 obtained in stage 1, we proceeded with stage 2 and
obtained a GAP fit providing test set errors of $\Gamma_E = 0.041$ eV/atom and
$\Gamma_F = 1.33$ eV/\AA{} by optimizing GAP HPs ($\delta = 0.36$~eV,
$\sigma_E = 0.05$~eV). 
aC4-2 with $\zeta = 2$ provided a fit with similar errors ($\Gamma_E = 0.049$
eV/atom and $\Gamma_F = 1.35$ eV/\AA{} using $\delta = 1.99$~eV and
$\sigma_E = 0.03$~eV) suggesting that, while large $\zeta$ values allow for better differentiation
of similar environments, they do not have a large effect on the quality of the
GAP fits. The above observations prove that our strategy 2 can be employed to
obtain reasonable GAPs from a pre-optimized SOAP for this complex a-C data set.
The force errors are very large when compared to the energy errors. It should be
possible to reduce the force errors by including force information to retrain the
GAPs with the above optimized HPs, which we will investigate further below.

\subsection{$\alpha$-Fe}
The optimal set of SOAP and GAP HPs for the alpha-Fe database obtained by
strategy 1 is given in Table~\ref{tab:HPOaC}. 
\begin{figure*}
    \centering
    \includegraphics[scale=0.4]{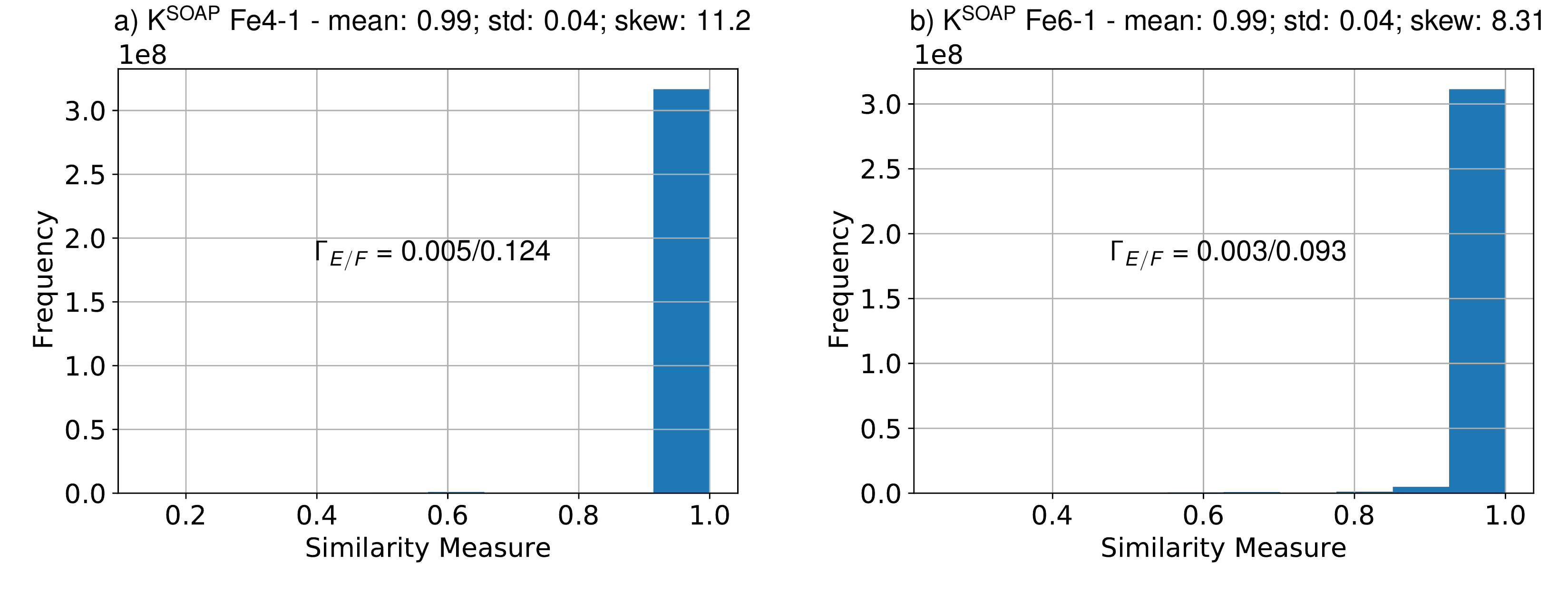}
    \caption{The SOAP kernel distribution of Fe4-1 (a) and Fe6-1 (b) obtained from HPs that minimized the test set RMSE.  }
    \label{fig:KFe64}
\end{figure*}
For the two optimal HP sets
Fe4-1 and Fe6-1, with differently sized SOAP, the energy and force errors
obtained for the test set are much smaller than for the a-C database. One reason
for this is that a-C database consists of many repulsive geometries with large
forces, compared to the $\alpha$-Fe database. The other reason is that bonding is
very complex in a-C, when compared to that in pure transition metals, and the amount of
cohesive energy per bond (related to the `spring constants' in traditional
force fields) is much higher. For instance, compare the bulk modulus and coordination
numbers of diamond ($\sim 450$~GPa and 4) and $\alpha$-Fe ($\sim 120$~GPa
and 8), which means that in C materials each bond stores about 7.5 times more 
energy than in Fe. Therefore, the \textit{relative} GAP errors are virtually the
same for a-C and $\alpha$-Fe databases. By contrast, the optimized SOAP kernel
distributions are much narrower for this database ($\text{std} = 0.04$) than for
the a-C database and do not span the entire kernel space,
as shown in Figure~\ref{fig:KFe64}.

Now we switch to stage 1 of strategy 2 to find a broad kernel distribution for
the small SOAP basis. We found a SOAP HP set, Fe4-2 ($\sigma_r = 0.24$~{\AA},
$\sigma_t = 0.17$~{\AA}, $\text{a}_r = 0.60$, $\text{a}_t = 0.0004$,
$\text{a} = 1.33$, $\text{cw} = 0.007$, $\text{re} = 2$,
$\text{rc}_\text{s} = 4.84$ and $\zeta = 8.98$), that gave a kernel distribution
that spanned the entire kernel space with a relatively larger std of 0.07 and a
skew of 8.5. Keeping this set of SOAP HPs fixed, we optimized the GAP HPs in
stage 2 ($\delta = 4.56$~eV, $\sigma_E = 0.005$~eV) 
to get the test set errors, $\Gamma_E = 4.5$ meV/atom and $\Gamma_F = 170$ meV/\AA.
Again, strategy 2 provided similar test set errors as that of strategy 1 for
the $\alpha$-Fe database.

\subsection{Training with Forces}
Using the optimal HP sets obtained in the last sections for the two complex data
sets, we retrained the GAPs on both energies and all force components (with
$\sigma_F = 0.02$~eV/{\AA}; we did not employ PSO to optimize this HP due to the
high computational cost and memory bottleneck) and the results are listed in
Table~\ref{tab:final}.
\begin{table}[]
    \centering
    \begin{tabular}{l c c c c}
    \hline\\
        HP set & \multicolumn{2}{c}{$\Gamma_E$ [eV/atom]} & \multicolumn{2}{c}{$\Gamma_F$ [eV/\AA]}\\
         & \multicolumn{2}{c}{GAP trained on} & \multicolumn{2}{c}{GAP trained on}\\
        & E only & E and F & E only & E and F\\
        \hline\\
         aC4-1&0.037&0.075&1.172&0.682\\
         aC4-2&0.041&0.098&1.330&0.755\\
         aC6-1&0.032&0.057&1.211&0.650\\
         Fe4-1&0.005&0.006&0.124&0.098\\
         Fe4-2&0.005&0.007&0.170&0.083\\
         Fe6-1&0.003&0.004&0.093&0.105\\
         \hline
    \end{tabular}
    \caption{Energy and force RMSEs of the GAPs trained only with reference energies (E only) compared to the GAPs retrained with both energies and forces (E and F).}
    \label{tab:final}
\end{table}
On the one hand,
we observe a significant reduction (roughly 50\%) in the force RMSE for the a-C GAPs
when compared to those trained on reference energies only. At the same time, the
energy RMSE for these GAPs has roughly doubled. Possible reasons for this observation
are that there are far more force data points than there are energies and we did not
optimize $\sigma_F$ with the other HPs. It is also possible to include only a fraction of
the force data for training, which would also reduce the training cost.
On the other hand, the force RMSEs for $\alpha$-Fe GAPs have improved only marginally.
However, the energy RMSEs have not increased much. One outlier is observed for the
$\alpha$-Fe HP set, Fe6-1, for which the retrained GAP gave a surprisingly larger force
error as compared to the energy only fit, but the difference is very small (12 meV/\AA).

\subsection{QM9 Dataset}
As a final test, we employed Strategy 1 to learn the potential energies of the open source QM9 dataset.~\cite{ramakrishnan2014quantum} The QM9 dataset consists of 130831 molecules made up of up to 5 elements (C, H, O, N, F) as listed in Table~\ref{tab:QM9comp}.

\begin{table}[h]
    \centering
    \begin{tabular}{l c c c c}
    \hline\\
         Elements& Mol. & Env. & Mol. 100 & Mol. 1000 \\
         \hline\\
CF	&	2	&	13	& &\\
CFH	&	96 	&	1491	&2 & 1 \\
CFHN	&	610	&	8536	&& 4\\
CFHNO	&	951	&	12170	&&4\\
CFHO	&	250	&	3540	&&\\
CFN	&	3	&	23	&&\\
CFNO	&	11	&	96	&&\\
CH	&	4890	&	106235	&4&31\\
CHN	&	13903	&	247483	&15&114\\
CHNO	&	64374	&	1099065	&47&516\\
CHO	&	45713	&	880348	&32&330\\
CN	&	4	&	27	&&\\
CNO	&	21	&	171	&&\\
HN	&	1	&	4	&&\\
HO	&	1	&	3	&&\\
NO	&	1	&	5	&&\\
Total	&	130831	&	2359210	&100&1000\\
 \hline
    \end{tabular}
    \caption{The composition of molecules in the QM9 dataset.}
    \label{tab:QM9comp}
\end{table}

In a recent benchmark study,~\cite{christensen2020fchl} the atomic descriptor
based spectrum of London and Axilrod–Teller–Muto (aSLATM) model was found to
provide the lowest MAE (between 210 meV/molecule and 230 meV/molecule) for the
entire dataset when only 100 molecules are used for training. We used PSO to
optimize a set of SOAP and GAP HPs so as to obtain an even lower MAE for the
entire dataset using specially chosen 100 molecules for training. The composition
of the training dataset is given in Table~\ref{tab:QM9comp}. The selection of
these 100 molecules was done by prescreening through many random combinations
of 100 molecules, training a GAP on those (without HPO) and then testing on a
small subset of the QM9 database (circa 10~\% of the entire
database). The combination of 100 molecules obtained this way was the one
chosen for HPO and then tested on the entire QM9 dataset.
We also performed a similar test using 1000 molecules (distributed as shown in
Table~\ref{tab:QM9comp}) in the dataset, although these 1000 molecules were
randomly chosen among those molecules that showed good learning power in the
100-molecule prescreening. Optimal selection of
molecular samples to construct such small datasets is far from trivial, and
there is current work in progress in our group in this regard. We will not
explore this issue further in this paper.

The optimized SOAP and GAP HPs for the two training datasets are listed in
Table~\ref{tab:HPOqm9}. For the optimization we fixed a small SOAP basis of
$n_{\text{max}} = l_{\text{max}} = 4$ which results in 1050 SOAP components per
atomic environment. If we had used a larger basis, say
$n_{\text{max}} = l_{\text{max}} = 8$, then we would have had to compute
4100 SOAP components per environment.
We also used a simple method to `compress' the SOAP descriptors, consisting
on retaining only some of the SOAP components (450 components in this case)
followed by renormalization of the SOAP vectors,
to speed up the calculation. The retained components are simply those
that incorporate expansion coefficients corresponding to the first radial
basis function (the one with the lowest associated singular value) in each species channel. For example,
for a single-species \texttt{soap\_turbo} descriptor, only components of the
form $p_{1n'l}$ are retained, and all others are dropped.
From our experience, this trivial compression does not affect the
quality of the GAPs too much. More information on compression schemes will be
provided in a future article. 

\begin{table}[h]
 \small
     \centering
     \begin{tabular}{l c c }
     \hline
        HP$^{\rm{SOAP}}$  & HP$_{100}$  & HP$_{1000}$\\
        \hline
        n$_{\rm{max}}$   &  4 & 4\\
        l$_{\rm{max}}$   &  4 & 4\\
        rc$_{\rm{h}}$  [\AA]  &  4.0 & 4.0 \\
        rc$_{\rm{s}}$  [\AA]  & 3.65 &  3.41 \\
        $\sigma_r$       & 0.48 &  0.33 \\
        $\sigma_t$       & 0.14 &  0.19\\
        a$_r$            & 0.44 &  0.001\\
        a$_t$            & 0.14 &  0.013\\
        a                & 1.97 &  1.98\\
        cw               & 0.67 &  0.40\\
        re               & 1   &    1 \\
        $\zeta$          & 2.23   &  4.43\\
        n$_{\rm{sparse}}$   & 2000 & 2000\\
        $\delta$         & 4.17 &  4.67\\
        $\sigma_E$  [meV/molecule]     & 0.001 & 0.02 \\\\
         MAE$_E$ [meV] & 198 & 64.5\\
        \hline
     \end{tabular}
     \caption{A list of SOAP and GAP HPs providing lowest MAE of energy for the full QM9 dataset.}
     \label{tab:HPOqm9}
 \end{table}
 
 We also fixed the SOAP hard cutoff to 4 \AA{} and the soft cutoff was allowed
 to be randomly chosen between 3 and 4 \AA{}. 2000 sparse environments are
 chosen by $CUR$ decomposition for the interpolation. All other initial
 parameters are chosen randomly within the range of values presented for earlier studies.
 
 At the end of the optimization, we obtained a MAE of 198 meV/molecule for the entire QM9 dataset using the 100 molecule training set and a MAE of 64.5 meV/molecule for the 1000 molecule training set. Our results are either similar to or better than the current best values in the literature.~\cite{christensen2020fchl} As a next step, We checked whether the HP set trained on a smaller training data provides lower errors when used with larger training data. When the HP$_{100}$ set is used to train on the 1000 molecule data set, the MAE went up to 84.3 meV/molecule. Similarly, when HP$_{1000}$ set is used to train on the 100 molecule data set, the MAE came to 313 meV/molecule. This suggests that HPs must be optimized for a given training data set.

%%%%%%%%%%%%%%%%%%%%%%%%%%%%%%%%%%%%%%%%%%%%%%%%%%%%%%%%%%%%%%%%%%%%%%%%%%%%%%%%%%%%%%%%%%%%%

\section{Conclusion}
We have presented two strategies to automatically optimize the hyper-parameters (HPs) for
ML-PES generation from a given data set, using SOAP descriptors and GAPs (although the
methodology is general and can be readily extended to other descriptor types and ML
architectures). A custom coded parallel particle swarm optimizer was employed to
stochastically find the optimal set of HPs in both strategies. This code is freely
available online for the community to make use of it. In strategy 1, both
SOAP and GAP HPs were optimized at the same time by minimizing the test
set RMSE. In strategy 2, the SOAP HPs are optimized first by maximizing the
standard deviation and minimizing the skewness of SOAP kernel distribution,
followed by the optimization of GAP HPs by minimizing the RMSE of the test set.
Strategy 2 was ultimately more efficient than strategy 1 since it did not require
many GAPs to be trained. Both strategies are validated using three data sets: a
simple toy C-dimer data set, together with more complex amorphous carbon and
$\alpha$-Fe data sets. Both strategies provided similar quality SOAPs and GAPs for
all of the tested data sets. More importantly, smaller SOAP basis was optimized to
provide similar quality of GAPs as that of larger SOAP basis for all data sets,
with the associated reduction in computational costs. Strategy 1 was then
used exclusively to train GAPs for predicting molecular energies of the QM9 dataset.
We demonstrated the ability of PSO to find optimal HPs for QM9 dataset using a smaller
SOAP basis and using as few as 100 molecules in the training set. Strategy 2 introduced in this
paper can be employed to efficiently optimize HPs of new and complex data sets without
much knowledge of the specific environments they contain. Techniques such as these are
needed to realise fully automated ML-PES generation.

\newpage
\begin{acknowledgement}
The authors acknowledge the funding for this research work provided by
the Academy of Finland under project numbers 321713 (S.~K.~N. \& M.~A.~C.),
310574, 329483 and 336304 (M.~A.~C.). Computational resources were provided
by CSC -- IT Center for Science.
\end{acknowledgement}

\providecommand{\latin}[1]{#1}
\makeatletter
\providecommand{\doi}
  {\begingroup\let\do\@makeother\dospecials
  \catcode`\{=1 \catcode`\}=2 \doi@aux}
\providecommand{\doi@aux}[1]{\endgroup\texttt{#1}}
\makeatother
\providecommand*\mcitethebibliography{\thebibliography}
\csname @ifundefined\endcsname{endmcitethebibliography}
  {\let\endmcitethebibliography\endthebibliography}{}


\begin{mcitethebibliography}{51}
\providecommand*\natexlab[1]{#1}
\providecommand*\mciteSetBstSublistMode[1]{}
\providecommand*\mciteSetBstMaxWidthForm[2]{}
\providecommand*\mciteBstWouldAddEndPuncttrue
  {\def\EndOfBibitem{\unskip.}}
\providecommand*\mciteBstWouldAddEndPunctfalse
  {\let\EndOfBibitem\relax}
\providecommand*\mciteSetBstMidEndSepPunct[3]{}
\providecommand*\mciteSetBstSublistLabelBeginEnd[3]{}
\providecommand*\EndOfBibitem{}
\mciteSetBstSublistMode{f}
\mciteSetBstMaxWidthForm{subitem}{(\alph{mcitesubitemcount})}
\mciteSetBstSublistLabelBeginEnd
  {\mcitemaxwidthsubitemform\space}
  {\relax}
  {\relax}

\bibitem[Allen and Tildesley(2017)Allen, and Tildesley]{allen2017computer}
Allen,~M.~P.; Tildesley,~D.~J. \emph{Computer simulation of liquids}; Oxford
  university press, 2017\relax
\mciteBstWouldAddEndPuncttrue
\mciteSetBstMidEndSepPunct{\mcitedefaultmidpunct}
{\mcitedefaultendpunct}{\mcitedefaultseppunct}\relax
\EndOfBibitem
\bibitem[Hollingsworth and Dror(2018)Hollingsworth, and
  Dror]{hollingsworth2018molecular}
Hollingsworth,~S.~A.; Dror,~R.~O. Molecular dynamics simulation for all.
  \emph{Neuron} \textbf{2018}, \emph{99}, 1129--1143\relax
\mciteBstWouldAddEndPuncttrue
\mciteSetBstMidEndSepPunct{\mcitedefaultmidpunct}
{\mcitedefaultendpunct}{\mcitedefaultseppunct}\relax
\EndOfBibitem
\bibitem[MacKerell~Jr(2004)]{mackerell2004empirical}
MacKerell~Jr,~A.~D. Empirical force fields for biological macromolecules:
  overview and issues. \emph{J. Comp. Chem.} \textbf{2004}, \emph{25},
  1584--1604\relax
\mciteBstWouldAddEndPuncttrue
\mciteSetBstMidEndSepPunct{\mcitedefaultmidpunct}
{\mcitedefaultendpunct}{\mcitedefaultseppunct}\relax
\EndOfBibitem
\bibitem[Marx and Hutter(2009)Marx, and Hutter]{marx2009ab}
Marx,~D.; Hutter,~J. \emph{Ab initio molecular dynamics: basic theory and
  advanced methods}; Cambridge University Press, 2009\relax
\mciteBstWouldAddEndPuncttrue
\mciteSetBstMidEndSepPunct{\mcitedefaultmidpunct}
{\mcitedefaultendpunct}{\mcitedefaultseppunct}\relax
\EndOfBibitem
\bibitem[Handley and Popelier(2010)Handley, and Popelier]{handley2010potential}
Handley,~C.~M.; Popelier,~P.~L. Potential energy surfaces fitted by artificial
  neural networks. \emph{J. Phys. Chem. A} \textbf{2010}, \emph{114},
  3371--3383\relax
\mciteBstWouldAddEndPuncttrue
\mciteSetBstMidEndSepPunct{\mcitedefaultmidpunct}
{\mcitedefaultendpunct}{\mcitedefaultseppunct}\relax
\EndOfBibitem
\bibitem[Behler(2016)]{behler2016perspective}
Behler,~J. Perspective: Machine learning potentials for atomistic simulations.
  \emph{J. Chem. Phys.} \textbf{2016}, \emph{145}, 170901\relax
\mciteBstWouldAddEndPuncttrue
\mciteSetBstMidEndSepPunct{\mcitedefaultmidpunct}
{\mcitedefaultendpunct}{\mcitedefaultseppunct}\relax
\EndOfBibitem
\bibitem[Schmitz \latin{et~al.}(2019)Schmitz, Godtliebsen, and
  Christiansen]{schmitz2019machine}
Schmitz,~G.; Godtliebsen,~I.~H.; Christiansen,~O. Machine learning for
  potential energy surfaces: An extensive database and assessment of methods.
  \emph{J. Chem. Phys.} \textbf{2019}, \emph{150}, 244113\relax
\mciteBstWouldAddEndPuncttrue
\mciteSetBstMidEndSepPunct{\mcitedefaultmidpunct}
{\mcitedefaultendpunct}{\mcitedefaultseppunct}\relax
\EndOfBibitem
\bibitem[Mueller \latin{et~al.}(2020)Mueller, Hernandez, and
  Wang]{mueller2020machine}
Mueller,~T.; Hernandez,~A.; Wang,~C. Machine learning for interatomic potential
  models. \emph{J. Chem. Phys.} \textbf{2020}, \emph{152}, 050902\relax
\mciteBstWouldAddEndPuncttrue
\mciteSetBstMidEndSepPunct{\mcitedefaultmidpunct}
{\mcitedefaultendpunct}{\mcitedefaultseppunct}\relax
\EndOfBibitem
\bibitem[Dral \latin{et~al.}(2020)Dral, Owens, Dral, and
  Cs{\'a}nyi]{dral2020hierarchical}
Dral,~P.~O.; Owens,~A.; Dral,~A.; Cs{\'a}nyi,~G. Hierarchical machine learning
  of potential energy surfaces. \emph{J. Chem. Phys.} \textbf{2020},
  \emph{152}, 204110\relax
\mciteBstWouldAddEndPuncttrue
\mciteSetBstMidEndSepPunct{\mcitedefaultmidpunct}
{\mcitedefaultendpunct}{\mcitedefaultseppunct}\relax
\EndOfBibitem
\bibitem[Deringer \latin{et~al.}(2019)Deringer, Caro, and
  Cs{\'a}nyi]{deringer_2019}
Deringer,~V.~L.; Caro,~M.~A.; Cs{\'a}nyi,~G. Machine Learning Interatomic
  Potentials as Emerging Tools for Materials Science. \emph{Adv. Mater.}
  \textbf{2019}, 1902765\relax
\mciteBstWouldAddEndPuncttrue
\mciteSetBstMidEndSepPunct{\mcitedefaultmidpunct}
{\mcitedefaultendpunct}{\mcitedefaultseppunct}\relax
\EndOfBibitem
\bibitem[Liu \latin{et~al.}(2010)Liu, Li, Ma, and Cheng]{liu2010advanced}
Liu,~C.; Li,~F.; Ma,~L.-P.; Cheng,~H.-M. Advanced materials for energy storage.
  \emph{Adv. Mater.} \textbf{2010}, \emph{22}, E28--E62\relax
\mciteBstWouldAddEndPuncttrue
\mciteSetBstMidEndSepPunct{\mcitedefaultmidpunct}
{\mcitedefaultendpunct}{\mcitedefaultseppunct}\relax
\EndOfBibitem
\bibitem[Butler \latin{et~al.}(2013)Butler, Hollen, Cao, Cui, Gupta,
  Guti{\'e}rrez, Heinz, Hong, Huang, Ismach, \latin{et~al.}
  others]{butler2013progress}
Butler,~S.~Z.; Hollen,~S.~M.; Cao,~L.; Cui,~Y.; Gupta,~J.~A.;
  Guti{\'e}rrez,~H.~R.; Heinz,~T.~F.; Hong,~S.~S.; Huang,~J.; Ismach,~A.~F.
  \latin{et~al.}  Progress, challenges, and opportunities in two-dimensional
  materials beyond graphene. \emph{ACS nano} \textbf{2013}, \emph{7},
  2898--2926\relax
\mciteBstWouldAddEndPuncttrue
\mciteSetBstMidEndSepPunct{\mcitedefaultmidpunct}
{\mcitedefaultendpunct}{\mcitedefaultseppunct}\relax
\EndOfBibitem
\bibitem[Fenton \latin{et~al.}(2018)Fenton, Olafson, Pillai, Mitchell, and
  Langer]{fenton2018advances}
Fenton,~O.~S.; Olafson,~K.~N.; Pillai,~P.~S.; Mitchell,~M.~J.; Langer,~R.
  Advances in biomaterials for drug delivery. \emph{Adv. Mater.} \textbf{2018},
  \emph{30}, 1705328\relax
\mciteBstWouldAddEndPuncttrue
\mciteSetBstMidEndSepPunct{\mcitedefaultmidpunct}
{\mcitedefaultendpunct}{\mcitedefaultseppunct}\relax
\EndOfBibitem
\bibitem[de~Pablo \latin{et~al.}(2019)de~Pablo, Jackson, Webb, Chen, Moore,
  Morgan, Jacobs, Pollock, Schlom, Toberer, \latin{et~al.} others]{de2019new}
de~Pablo,~J.~J.; Jackson,~N.~E.; Webb,~M.~A.; Chen,~L.-Q.; Moore,~J.~E.;
  Morgan,~D.; Jacobs,~R.; Pollock,~T.; Schlom,~D.~G.; Toberer,~E.~S.
  \latin{et~al.}  New frontiers for the materials genome initiative. \emph{Npj
  Comput. Mater} \textbf{2019}, \emph{5}, 41\relax
\mciteBstWouldAddEndPuncttrue
\mciteSetBstMidEndSepPunct{\mcitedefaultmidpunct}
{\mcitedefaultendpunct}{\mcitedefaultseppunct}\relax
\EndOfBibitem
\bibitem[Schmidt \latin{et~al.}(2019)Schmidt, Marques, Botti, and
  Marques]{schmidt2019recent}
Schmidt,~J.; Marques,~M.~R.; Botti,~S.; Marques,~M.~A. Recent advances and
  applications of machine learning in solid-state materials science. \emph{Npj
  Comput. Mater} \textbf{2019}, \emph{5}, 1--36\relax
\mciteBstWouldAddEndPuncttrue
\mciteSetBstMidEndSepPunct{\mcitedefaultmidpunct}
{\mcitedefaultendpunct}{\mcitedefaultseppunct}\relax
\EndOfBibitem
\bibitem[Meredig(2019)]{2019-challenges-Bryce}
Meredig,~B. Five High-Impact Research Areas in Machine Learning for Materials
  Science. \emph{Chem. Mater.} \textbf{2019}, \emph{31}, 9579--9581\relax
\mciteBstWouldAddEndPuncttrue
\mciteSetBstMidEndSepPunct{\mcitedefaultmidpunct}
{\mcitedefaultendpunct}{\mcitedefaultseppunct}\relax
\EndOfBibitem
\bibitem[Morgan and Jacobs(2020)Morgan, and Jacobs]{morgan2020opportunities}
Morgan,~D.; Jacobs,~R. Opportunities and Challenges for Machine Learning in
  Materials Science. \emph{Annu. Rev. Mater.} \textbf{2020}, \emph{50}\relax
\mciteBstWouldAddEndPuncttrue
\mciteSetBstMidEndSepPunct{\mcitedefaultmidpunct}
{\mcitedefaultendpunct}{\mcitedefaultseppunct}\relax
\EndOfBibitem
\bibitem[Kohn and Sham(1965)Kohn, and Sham]{kohn1965self}
Kohn,~W.; Sham,~L.~J. Self-consistent equations including exchange and
  correlation effects. \emph{Phys. Rev.} \textbf{1965}, \emph{140}, A1133\relax
\mciteBstWouldAddEndPuncttrue
\mciteSetBstMidEndSepPunct{\mcitedefaultmidpunct}
{\mcitedefaultendpunct}{\mcitedefaultseppunct}\relax
\EndOfBibitem
\bibitem[Schran \latin{et~al.}(2018)Schran, Uhl, Behler, and
  Marx]{schran2018high}
Schran,~C.; Uhl,~F.; Behler,~J.; Marx,~D. High-dimensional neural network
  potentials for solvation: The case of protonated water clusters in helium.
  \emph{J. Chem. Phys.} \textbf{2018}, \emph{148}, 102310\relax
\mciteBstWouldAddEndPuncttrue
\mciteSetBstMidEndSepPunct{\mcitedefaultmidpunct}
{\mcitedefaultendpunct}{\mcitedefaultseppunct}\relax
\EndOfBibitem
\bibitem[Chmiela \latin{et~al.}(2018)Chmiela, Sauceda, Müller, and
  Tkatchenko]{chmiela2018}
Chmiela,~S.; Sauceda,~H.~E.; Müller,~K.-R.; Tkatchenko,~A. Towards exact
  molecular dynamics simulations with machine-learned force fields. \emph{Nat.
  Commun.} \textbf{2018}, \emph{9}, 3887\relax
\mciteBstWouldAddEndPuncttrue
\mciteSetBstMidEndSepPunct{\mcitedefaultmidpunct}
{\mcitedefaultendpunct}{\mcitedefaultseppunct}\relax
\EndOfBibitem
\bibitem[Rupp \latin{et~al.}(2012)Rupp, Tkatchenko, M{\"u}ller, and
  Von~Lilienfeld]{rupp2012fast}
Rupp,~M.; Tkatchenko,~A.; M{\"u}ller,~K.-R.; Von~Lilienfeld,~O.~A. Fast and
  accurate modeling of molecular atomization energies with machine learning.
  \emph{Phys. Rev. Lett.} \textbf{2012}, \emph{108}, 058301\relax
\mciteBstWouldAddEndPuncttrue
\mciteSetBstMidEndSepPunct{\mcitedefaultmidpunct}
{\mcitedefaultendpunct}{\mcitedefaultseppunct}\relax
\EndOfBibitem
\bibitem[Huo and Rupp(2017)Huo, and Rupp]{huo2017unified}
Huo,~H.; Rupp,~M. Unified representation of molecules and crystals for machine
  learning. \emph{arXiv preprint arXiv:1704.06439} \textbf{2017}, \relax
\mciteBstWouldAddEndPunctfalse
\mciteSetBstMidEndSepPunct{\mcitedefaultmidpunct}
{}{\mcitedefaultseppunct}\relax
\EndOfBibitem
\bibitem[Himanen \latin{et~al.}(2020)Himanen, J{\"a}ger, Morooka, Canova,
  Ranawat, Gao, Rinke, and Foster]{himanen2020dscribe}
Himanen,~L.; J{\"a}ger,~M.~O.; Morooka,~E.~V.; Canova,~F.~F.; Ranawat,~Y.~S.;
  Gao,~D.~Z.; Rinke,~P.; Foster,~A.~S. DScribe: Library of descriptors for
  machine learning in materials science. \emph{Comput. Phys. Commun.}
  \textbf{2020}, \emph{247}, 106949\relax
\mciteBstWouldAddEndPuncttrue
\mciteSetBstMidEndSepPunct{\mcitedefaultmidpunct}
{\mcitedefaultendpunct}{\mcitedefaultseppunct}\relax
\EndOfBibitem
\bibitem[Christensen \latin{et~al.}(2020)Christensen, Bratholm, Faber, and
  Anatole~von Lilienfeld]{christensen2020fchl}
Christensen,~A.~S.; Bratholm,~L.~A.; Faber,~F.~A.; Anatole~von Lilienfeld,~O.
  FCHL revisited: Faster and more accurate quantum machine learning. \emph{J.
  Chem. Phys.} \textbf{2020}, \emph{152}, 044107\relax
\mciteBstWouldAddEndPuncttrue
\mciteSetBstMidEndSepPunct{\mcitedefaultmidpunct}
{\mcitedefaultendpunct}{\mcitedefaultseppunct}\relax
\EndOfBibitem
\bibitem[Bart{\'o}k \latin{et~al.}(2013)Bart{\'o}k, Kondor, and
  Cs{\'a}nyi]{bartok2013representing}
Bart{\'o}k,~A.~P.; Kondor,~R.; Cs{\'a}nyi,~G. On representing chemical
  environments. \emph{Phys. Rev. B} \textbf{2013}, \emph{87}, 184115\relax
\mciteBstWouldAddEndPuncttrue
\mciteSetBstMidEndSepPunct{\mcitedefaultmidpunct}
{\mcitedefaultendpunct}{\mcitedefaultseppunct}\relax
\EndOfBibitem
\bibitem[Caro(2019)]{caro2019optimizing}
Caro,~M.~A. Optimizing many-body atomic descriptors for enhanced computational
  performance of machine learning based interatomic potentials. \emph{Phys.
  Rev. B} \textbf{2019}, \emph{100}, 024112\relax
\mciteBstWouldAddEndPuncttrue
\mciteSetBstMidEndSepPunct{\mcitedefaultmidpunct}
{\mcitedefaultendpunct}{\mcitedefaultseppunct}\relax
\EndOfBibitem
\bibitem[Behler(2011)]{behler2011atom}
Behler,~J. Atom-centered symmetry functions for constructing high-dimensional
  neural network potentials. \emph{J. Chem. Phys.} \textbf{2011}, \emph{134},
  074106\relax
\mciteBstWouldAddEndPuncttrue
\mciteSetBstMidEndSepPunct{\mcitedefaultmidpunct}
{\mcitedefaultendpunct}{\mcitedefaultseppunct}\relax
\EndOfBibitem
\bibitem[Willatt \latin{et~al.}(2019)Willatt, Musil, and
  Ceriotti]{willatt2019atom}
Willatt,~M.~J.; Musil,~F.; Ceriotti,~M. Atom-density representations for
  machine learning. \emph{J. Chem. Phys.} \textbf{2019}, \emph{150},
  154110\relax
\mciteBstWouldAddEndPuncttrue
\mciteSetBstMidEndSepPunct{\mcitedefaultmidpunct}
{\mcitedefaultendpunct}{\mcitedefaultseppunct}\relax
\EndOfBibitem
\bibitem[Imbalzano \latin{et~al.}(2018)Imbalzano, Anelli, Giofr{\'e}, Klees,
  Behler, and Ceriotti]{imbalzano2018automatic}
Imbalzano,~G.; Anelli,~A.; Giofr{\'e},~D.; Klees,~S.; Behler,~J.; Ceriotti,~M.
  Automatic selection of atomic fingerprints and reference configurations for
  machine-learning potentials. \emph{J. Chem. Phys.} \textbf{2018}, \emph{148},
  241730\relax
\mciteBstWouldAddEndPuncttrue
\mciteSetBstMidEndSepPunct{\mcitedefaultmidpunct}
{\mcitedefaultendpunct}{\mcitedefaultseppunct}\relax
\EndOfBibitem
\bibitem[Sch{\"u}tt \latin{et~al.}(2017)Sch{\"u}tt, Arbabzadah, Chmiela,
  M{\"u}ller, and Tkatchenko]{schutt2017quantum}
Sch{\"u}tt,~K.~T.; Arbabzadah,~F.; Chmiela,~S.; M{\"u}ller,~K.~R.;
  Tkatchenko,~A. Quantum-Chemical Insights From Deep Tensor Neural Networks.
  \emph{Nat. Commun.} \textbf{2017}, \emph{8}, 1--8\relax
\mciteBstWouldAddEndPuncttrue
\mciteSetBstMidEndSepPunct{\mcitedefaultmidpunct}
{\mcitedefaultendpunct}{\mcitedefaultseppunct}\relax
\EndOfBibitem
\bibitem[Behler and Parrinello(2007)Behler, and
  Parrinello]{behler2007generalized}
Behler,~J.; Parrinello,~M. Generalized neural-network representation of
  high-dimensional potential-energy surfaces. \emph{Phys. Rev. Lett.}
  \textbf{2007}, \emph{98}, 146401\relax
\mciteBstWouldAddEndPuncttrue
\mciteSetBstMidEndSepPunct{\mcitedefaultmidpunct}
{\mcitedefaultendpunct}{\mcitedefaultseppunct}\relax
\EndOfBibitem
\bibitem[Bart{\'o}k \latin{et~al.}(2010)Bart{\'o}k, Payne, Kondor, and
  Cs{\'a}nyi]{bartok2010gaussian}
Bart{\'o}k,~A.~P.; Payne,~M.~C.; Kondor,~R.; Cs{\'a}nyi,~G. Gaussian
  approximation potentials: The accuracy of quantum mechanics, without the
  electrons. \emph{Phys. Rev. Lett.} \textbf{2010}, \emph{104}, 136403\relax
\mciteBstWouldAddEndPuncttrue
\mciteSetBstMidEndSepPunct{\mcitedefaultmidpunct}
{\mcitedefaultendpunct}{\mcitedefaultseppunct}\relax
\EndOfBibitem
\bibitem[Bart{\'o}k and Cs{\'a}nyi(2015)Bart{\'o}k, and
  Cs{\'a}nyi]{bartok2015g}
Bart{\'o}k,~A.~P.; Cs{\'a}nyi,~G. Gaussian approximation potentials: A brief
  tutorial introduction. \emph{Int. J. Quantum Chem.} \textbf{2015},
  \emph{115}, 1051--1057\relax
\mciteBstWouldAddEndPuncttrue
\mciteSetBstMidEndSepPunct{\mcitedefaultmidpunct}
{\mcitedefaultendpunct}{\mcitedefaultseppunct}\relax
\EndOfBibitem
\bibitem[Bernstein \latin{et~al.}(2019)Bernstein, Cs{\'a}nyi, and
  Deringer]{bernstein2019novo}
Bernstein,~N.; Cs{\'a}nyi,~G.; Deringer,~V.~L. De novo exploration and
  self-guided learning of potential-energy surfaces. \emph{npj Comput. Mater.}
  \textbf{2019}, \emph{5}, 1\relax
\mciteBstWouldAddEndPuncttrue
\mciteSetBstMidEndSepPunct{\mcitedefaultmidpunct}
{\mcitedefaultendpunct}{\mcitedefaultseppunct}\relax
\EndOfBibitem
\bibitem[Bergstra \latin{et~al.}(2015)Bergstra, Komer, Eliasmith, Yamins, and
  Cox]{bergstra2015hyperopt}
Bergstra,~J.; Komer,~B.; Eliasmith,~C.; Yamins,~D.; Cox,~D.~D. Hyperopt: a
  python library for model selection and hyperparameter optimization.
  \emph{Comput. Sci. Discov.} \textbf{2015}, \emph{8}, 014008\relax
\mciteBstWouldAddEndPuncttrue
\mciteSetBstMidEndSepPunct{\mcitedefaultmidpunct}
{\mcitedefaultendpunct}{\mcitedefaultseppunct}\relax
\EndOfBibitem
\bibitem[Abbott \latin{et~al.}(2019)Abbott, Turney, Zhang, Smith, Altarawy, and
  Schaefer]{doi:10.1021/acs.jctc.9b00312}
Abbott,~A.~S.; Turney,~J.~M.; Zhang,~B.; Smith,~D. G.~A.; Altarawy,~D.;
  Schaefer,~H.~F. PES-Learn: An Open-Source Software Package for the Automated
  Generation of Machine Learning Models of Molecular Potential Energy Surfaces.
  \emph{J. Chem. Theory Comput.} \textbf{2019}, \emph{15}, 4386--4398, PMID:
  31283237\relax
\mciteBstWouldAddEndPuncttrue
\mciteSetBstMidEndSepPunct{\mcitedefaultmidpunct}
{\mcitedefaultendpunct}{\mcitedefaultseppunct}\relax
\EndOfBibitem
\bibitem[Collet and Rennard(2008)Collet, and Rennard]{collet2008stochastic}
Collet,~P.; Rennard,~J.-P. \emph{Intelligent information technologies:
  Concepts, methodologies, tools, and applications}; IGI Global, 2008; pp
  1121--1137\relax
\mciteBstWouldAddEndPuncttrue
\mciteSetBstMidEndSepPunct{\mcitedefaultmidpunct}
{\mcitedefaultendpunct}{\mcitedefaultseppunct}\relax
\EndOfBibitem
\bibitem[Feurer and Hutter(2019)Feurer, and Hutter]{Feurer2019}
Feurer,~M.; Hutter,~F. In \emph{Automated Machine Learning: Methods, Systems,
  Challenges}; Hutter,~F., Kotthoff,~L., Vanschoren,~J., Eds.; Springer
  International Publishing: Cham, 2019; pp 3--33\relax
\mciteBstWouldAddEndPuncttrue
\mciteSetBstMidEndSepPunct{\mcitedefaultmidpunct}
{\mcitedefaultendpunct}{\mcitedefaultseppunct}\relax
\EndOfBibitem
\bibitem[Cawley \latin{et~al.}(2005)Cawley, Talbot, and
  Chapelle]{cawley2005estimating}
Cawley,~G.~C.; Talbot,~N.~L.~C.; Chapelle,~O. Estimating predictive variances
  with kernel ridge regression. Machine Learning Challenges Workshop. 2005;
  p~56\relax
\mciteBstWouldAddEndPuncttrue
\mciteSetBstMidEndSepPunct{\mcitedefaultmidpunct}
{\mcitedefaultendpunct}{\mcitedefaultseppunct}\relax
\EndOfBibitem
\bibitem[Deringer and Cs{\'a}nyi(2017)Deringer, and
  Cs{\'a}nyi]{deringer2017machine}
Deringer,~V.~L.; Cs{\'a}nyi,~G. Machine learning based interatomic potential
  for amorphous carbon. \emph{Phys. Rev. B} \textbf{2017}, \emph{95},
  094203\relax
\mciteBstWouldAddEndPuncttrue
\mciteSetBstMidEndSepPunct{\mcitedefaultmidpunct}
{\mcitedefaultendpunct}{\mcitedefaultseppunct}\relax
\EndOfBibitem
\bibitem[Anderson and Whitcomb(2000)Anderson, and Whitcomb]{anderson2000design}
Anderson,~M.~J.; Whitcomb,~P.~J. Design of experiments. \emph{Kirk-Othmer
  Encyclopedia of Chemical Technology} \textbf{2000}, 1--22\relax
\mciteBstWouldAddEndPuncttrue
\mciteSetBstMidEndSepPunct{\mcitedefaultmidpunct}
{\mcitedefaultendpunct}{\mcitedefaultseppunct}\relax
\EndOfBibitem
\bibitem[Hedayat \latin{et~al.}(2012)Hedayat, Sloane, and
  Stufken]{hedayat2012orthogonal}
Hedayat,~A.~S.; Sloane,~N. J.~A.; Stufken,~J. \emph{Orthogonal arrays: theory
  and applications}; Springer Science \& Business Media, 2012\relax
\mciteBstWouldAddEndPuncttrue
\mciteSetBstMidEndSepPunct{\mcitedefaultmidpunct}
{\mcitedefaultendpunct}{\mcitedefaultseppunct}\relax
\EndOfBibitem
\bibitem[Sobol'(1967)]{SOBOL196786}
Sobol',~I. On the distribution of points in a cube and the approximate
  evaluation of integrals. \emph{USSR Comput. Math. \& Math. Phys.}
  \textbf{1967}, \emph{7}, 86 -- 112\relax
\mciteBstWouldAddEndPuncttrue
\mciteSetBstMidEndSepPunct{\mcitedefaultmidpunct}
{\mcitedefaultendpunct}{\mcitedefaultseppunct}\relax
\EndOfBibitem
\bibitem[Kondati~Natarajan(2020)]{sureshDOE}
Kondati~Natarajan,~S. DOE-offline-optimization.
  \url{https://github.com/suresh0807/DOE-offline-optimization.git}, 2020\relax
\mciteBstWouldAddEndPuncttrue
\mciteSetBstMidEndSepPunct{\mcitedefaultmidpunct}
{\mcitedefaultendpunct}{\mcitedefaultseppunct}\relax
\EndOfBibitem
\bibitem[Snoek \latin{et~al.}(2012)Snoek, Larochelle, and
  Adams]{snoek2012practical}
Snoek,~J.; Larochelle,~H.; Adams,~R.~P. Practical bayesian optimization of
  machine learning algorithms. Advances in neural information processing
  systems. 2012; pp 2951--2959\relax
\mciteBstWouldAddEndPuncttrue
\mciteSetBstMidEndSepPunct{\mcitedefaultmidpunct}
{\mcitedefaultendpunct}{\mcitedefaultseppunct}\relax
\EndOfBibitem
\bibitem[qui()]{quip}
QUIP. \url{https://libatoms.github.io}, Accessed: 2020-08-17\relax
\mciteBstWouldAddEndPuncttrue
\mciteSetBstMidEndSepPunct{\mcitedefaultmidpunct}
{\mcitedefaultendpunct}{\mcitedefaultseppunct}\relax
\EndOfBibitem
\bibitem[tur()]{turbogap}
TurboGAP. \url{https://turbogap.fi/wiki/index.php/Main_Page}, Accessed:
  2020-08-17\relax
\mciteBstWouldAddEndPuncttrue
\mciteSetBstMidEndSepPunct{\mcitedefaultmidpunct}
{\mcitedefaultendpunct}{\mcitedefaultseppunct}\relax
\EndOfBibitem
\bibitem[Dragoni \latin{et~al.}(2017)Dragoni, Daff, Cs{\'a}nyi, and
  Marzari]{Feds}
Dragoni,~D.; Daff,~T.~D.; Cs{\'a}nyi,~G.; Marzari,~N. {Gaussian Approximation
  Potentials for iron from extended first-principles database (Data Download),
  Materials Cloud Archive 2017.0006/v2 (2017)}.
  \url{https://archive.materialscloud.org/record/2017.0006/v2}, 2017\relax
\mciteBstWouldAddEndPuncttrue
\mciteSetBstMidEndSepPunct{\mcitedefaultmidpunct}
{\mcitedefaultendpunct}{\mcitedefaultseppunct}\relax
\EndOfBibitem
\bibitem[Mahoney and Drineas(2009)Mahoney, and Drineas]{mahoney2009cur}
Mahoney,~M.~W.; Drineas,~P. CUR matrix decompositions for improved data
  analysis. \emph{Proc. Natl. Acad.} \textbf{2009}, \emph{106}, 697--702\relax
\mciteBstWouldAddEndPuncttrue
\mciteSetBstMidEndSepPunct{\mcitedefaultmidpunct}
{\mcitedefaultendpunct}{\mcitedefaultseppunct}\relax
\EndOfBibitem
\bibitem[Ramakrishnan \latin{et~al.}(2014)Ramakrishnan, Dral, Rupp, and von
  Lilienfeld]{ramakrishnan2014quantum}
Ramakrishnan,~R.; Dral,~P.~O.; Rupp,~M.; von Lilienfeld,~O.~A. Quantum
  chemistry structures and properties of 134 kilo molecules. \emph{Scientific
  Data} \textbf{2014}, \emph{1}\relax
\mciteBstWouldAddEndPuncttrue
\mciteSetBstMidEndSepPunct{\mcitedefaultmidpunct}
{\mcitedefaultendpunct}{\mcitedefaultseppunct}\relax
\EndOfBibitem
\end{mcitethebibliography}
\end{document}